\documentclass[useAMS,usenatbib]{mn2e}
\usepackage{times}
\usepackage{graphicx}
\usepackage{amsfonts}
\usepackage{url} 
\def\ltsima{$\; \buildrel < \over \sim \;$}
\def\lsim{\lower.5ex\hbox{\ltsima}}
\def\gtsima{$\; \buildrel > \over \sim \;$}
\def\gsim{\lower.5ex\hbox{\gtsima}}
\def\ga{\mathrel{\hbox{\rlap{\hbox{\lower4pt\hbox{$\sim$}}}\hbox{$>$}}}}
\def\la{\mathrel{\hbox{\rlap{\hbox{\lower4pt\hbox{$\sim$}}}\hbox{$<$}}}}

\newcommand{\cMpch}{~h^{-1}~\mbox{comoving Mpc}}
\newcommand{\Msunh}{~h^{-1}~\mbox{M}_{\odot}}

\newcommand{\cmbfast}{{\sc cmbfast}}

\title[On the spin-temperature evolution during the EoR]{On the spin-temperature evolution during the epoch of reionization}

\author[Thomas et al.,]{Rajat M.
Thomas\footnote{E-mail:rajat.thomas@ipmu.jp}, 
Saleem Zaroubi\footnote{saleem@astro.rug.nl},
\\
Kapteyn Astronomical Institute, Landleven 12,Groningen 9747 AD, The
Netherlands \\ 
Max-Planck-Institut fuer Astrophysik,
Karl-Schwarzschild-Strasse 1, D-85748 Garching b. Muenchen, Germany}

\author[Thomas et al., ]{ Rajat M. Thomas$^{1,2,3}$\thanks{E-mail:
thomas@cita.utoronto.ca}, Saleem Zaroubi$^{1,4}$  
\\ $^{1}$Kapteyn Astronomical Institute,
University of Groningen, P.O. Box 800, 9700 AV Groningen, the
Netherlands.\\$^{2}$Institute for the Mathematics and Physics of the
Universe (IPMU), The University of Tokyo, Chiba, 277-8582, Japan.
\\$^{3}$CITA, University of Toronto, 60. St. George Street, M5S 3H8,
Toronto, ON, Canada.\\$^{4}$ Physics Department, Technion, Haifa 32000, Israel.
 }

\begin{document}

\date{}

\pagerange{\pageref{firstpage}--\pageref{lastpage}} \pubyear{2009}

\maketitle

\label{firstpage}

\begin{abstract}

  \noindent Simulations estimating the brightness temperature ($\delta
  \mathrm{T_b}$) of the redshifted 21-cm from the epoch of
  reionization (EoR) often assume that the spin temperature
  ($\mathrm{T_s}$) is decoupled from the background CMB temperature
  and is much larger than it, i.e., $\mathrm{T_s} \gg
  \mathrm{T_{CMB}}$. Although a valid assumption towards the latter
  stages of the reionization process, it does not necessarily hold at
  the earlier epochs. Violation of this assumption will lead to
  fluctuations in $\delta \mathrm{T_b}$ that are neither driven by
  density fluctuations nor by $\mathrm{H_{II}}$ regions. Therefore, it
  is vital to calculate the spin temperature self-consistently by
  treating the Ly$\alpha$ and collisional coupling of
  $\mathrm{T_s}$ to the kinetic temperature, $\mathrm{T_k}$. In this
  paper we develop an extension to the \textsc{bears} algorithm,
  originally developed to model reionization history, to include these
  coupling effects. Here we simulate the effect in ionization and
  heating for three models in which the reionization is driven by
  stars, miniqsos or a mixture of both.  We also perform a number of
  statistical tests to quantify the imprint of the self-consistent
  inclusion of the spin temperature decoupling from the CMB. We find
  that the evolution of the spin temperature has an impact on the
  measured signal specially at redshifts higher than 10 and such
  evolution should be taken into account when one attempts to
  interpret the observational data.

\end{abstract}

\begin{keywords} Physical data processes: radiative transfer --
  cosmology: theory -- observation -- diffuse radiation -- radio
  lines: general.  \end{keywords}

\section{Introduction}
\label{sec:intro}
Physical processes that occur during reionization are numerous and
complex. Nevertheless, ionization of neutral gas (hydrogen \& helium)
and heating of the inter-galactic medium (IGM) can be considered the two
primary influences of radiating objects during reionization.

Currently, the most promising ``direct'' probe of reionization is the
redshifted 21-cm radiation emanating from neutral hydrogen during the
epoch of reionization (EoR), which are to be measured using upcoming
telescopes like
LOFAR\footnote{\url{www.lofar.org}}'\footnote{\url{www.astro.rug.nl/~LofarEoR}},
MWA\footnote{\url{http://www.mwatelescope.org/}},
PAPER\footnote{\url{http://astro.berkeley.edu/~dbacker/eor/}} and
21CMA\footnote{\url{http://21cma.bao.ac.cn/}}. The intensity of the
observed 21-cm radiation depends on the ratio between the number
density of electrons in the hyperfine states in the ground state of a
neutral hydrogen atom. This ratio is normally expressed in terms of
the so-called 21-cm spin temperature, $\mathrm{T_s}$.  At the onset of
the formation of the first reionizing objects the spin temperature is
equal to the CMB temperature since at these redshifts the ratio
between excited and ground hyperfine state electrons is completely
determined by the CMB. However, as the number of ionizing sources
increases, $\mathrm{T_s}$ starts departing from $\mathrm{T_{CMB}}$;
slowly at the beginning, then rapidly approaching values larger than
$\mathrm{T_{CMB}}$.  This evolution is typically ignored in most
previous studies of reionization which assumes $\mathrm{T_s}
\gg \mathrm{T_{CMB}}$ at all times
\citep{mellema06,abel00,abel02,bromm02,yoshida03}.
  
Recently, \citet{baek09} have relaxed this assumption on
$\mathrm{T_s}$ at the dawn of reionization and explored its impact on
the brightness temperature. They found a considerable considerable
deviation from assuming $\mathrm{T_s} \gg \mathrm{T_{CMB}}$ at the
beginning of reionization. Towards the end of reionization though,
this assumption holds ground.  But, in order to track the evolution of
$\mathrm{T_s}$ accurately, like in \cite{baek09}, it is necessary to
perform a detailed 3-D Ly$\alpha$ radiative transfer calculation. The
Ly$\alpha$ photons undergo a large number ($\sim 10^5$) of scatterings
even in a marginally neutral medium before it is sufficiently off
line-centre to ``free stream''. The scattering angle after each
encounter is completely random and therefore the radiative transfer is
often done in a Monte Carlo sense \citep{tasitsiomi06,
  baek09,pierleoni09,laursen09} to capture this random nature of
Ly$\alpha$ scatterings.

 Unfortunately these Monte Carlo radiative transfer schemes are
 computationally very expensive, especially if we need to simulate
 large fields of view necessary to generate mock data sets for next
 generation radio telescopes. In order to circumvent the need to
 perform such computer-intensive calculations to obtain
 $\mathrm{T_s}$, we develop an algorithm along the lines of
 \textsc{bears} \citep{thomas09} as an approximation. In this paper we
 present an algorithm that follows the decoupling of $\mathrm{T_s}$
 from $\mathrm{T_{CMB}}$ owing to Ly$\alpha$ photons, which couples
 the spin temperature to the colour/kinetic temperature via the
 Wouthuysen-Field effect \citep{wouthuysen52}.  Collisional excitation
 and heating caused by secondary electrons resulting from hard X-ray
 radiation are also included. The dominant source of
   Ly$\alpha$ flux is the background created by the redshifting of
   photons in the Lyman band into Ly$\alpha$. These photons are
   blueward of Ly$\alpha$ and is injected into Ly$\alpha$ at some
   distance away from the source.

 The amount of intrinsic Ly$\alpha$, ionizing and ``heating'' photons
 is a function of the source spectral energy distribution (SED). Thus
 the evolution of the spin temperature critically depends on the
 source of reionization.  Different reionization sources manifest
 themselves by influencing the IGM in markedly different ways. For
 example, deficiency of hard photons in the SEDs of ``first stars'',
 limit the extent to which they heat the IGM
 \citep{thomas08,zaroubi07,zaroubi05}, while miniquasars (or miniqsos,
 characterized by central black hole masses less than a million
 solar), abundant in X-ray photons, cause considerable heating
 \citep{madau97,ricotti04a,nusser05,furlanetto04b,wyithe04,
   furlanetto02,thomas08,zaroubi07}. Ionization profiles similarly
 have their characteristic source-dependent behavior.

 Although the question on which sources did the bulk of the
 reionization is up for debate, it is conceivable from observations of
 the local Universe up to redshifts around 6.5, that sources of
 reionization could have been a mixture of both stellar and quasar
 kinds (their respective roles again are uncertain). Implementing
 radiative transfer that include both ionizing and hard X-ray photons
 has been difficult and as a result most 3-D radiative transfer
 schemes restrict themselves to ionization due to stars
 \citep{gnedin01,ciardi01,ritzerveld03,susa06,razoumov05,
   nakamoto01,whalen06,rijkhorst06,mellema06,zahn07,mesinger07,pawlik08}.
 In \cite{ricotti04a}, a ``semi'' hybrid model of stars and miniqsos,
 like the one hinted above, has been used albeit in sequential order
 instead of a simultaneous implementation. That is, pre-ionization due
 to miniqsos was invoked between $7 \leq z \leq 20$, after which,
 stars reionize the Universe at redshift 7. We in this paper would
 like to address the issue of simulating the propagation of both the
 UV and hard X-ray photons, exactly in 1-D and as approximation in
 3-D.

The focus of this paper is therefore to introduce the algorithm that
is used to implement IGM heating in \textsc{bears} along with the
procedure to estimate the spin temperature of the IGM. As an
application of this technique we explore the effects of heating due to
miniqsos, stars and, for the first time, a mixed ``hybrid
population''. Subsequently, we provide quantitative and qualitative
analysis of the differences in the 21-cm EoR signal with and without
the usual assumption of $\mathrm{T_s}$ being always decoupled from
$\mathrm{T_{CMB}}$.

The paper is organized as follows; \S\ref{sec:simulations} describes
briefly the N-body and 1-D radiative transfer codes used. In
\S\ref{sec:bearheat} we describe the adaptation of \textsc{bears} to
include $\mathrm{T_k}$, followed by the calculation of the
$\mathrm{T_s}$ and $\delta \mathrm{T_b}$ within the simulation
box. \textsc{bears} is then applied to three different scenarios of
reionization in \S\ref{sec:applications}, \emph{viz}., (1) the primary
source being stars, (2) miniqsos and (3) a hybrid population of stars
and miniqsos. Subsequently, observational cubes of $\delta
\mathrm{T_b}$ are generated for each of these scenarios and its
properties discussed. In \S\ref{sec:stats} we provide statistics on
the simulated boxes and interpret the finding mainly from the point of
view of the differences in $\delta \mathrm{T_b}$ with and without the
usual assumption that $\mathrm{T_s}$ is decoupled from $
\mathrm{T_{CMB}}$. We also compare our work to that of \cite{santos08}
in the same section. Conclusions and discussions of the results are
presented in \S\ref{sec:conclusion}, along with a mention of a few
topics that can be addressed using the data set simulated in this
paper.

\section{Simulations}
\label{sec:simulations}

In this section we briefly mention the dark-matter-only N-body
simulations employed, the 1-D radiative transfer code developed by
 \citet{thomas08} and provide an overview of the \textsc{bears}
 algorithm designed in \cite{thomas09}.

\subsection{N-body: dark matter only}
The dark matter only N-body simulations are the same as in
\citet{thomas09} and was performed by Joop Schaye and Andreas Pawlik
at the Leiden observatory.  The N-body/TreePM/SPH code \textsc{gadget}
\citep{springel05} was used to perform a dark matter (DM) cosmological
simulation containing $512^3$ particles in a box of size $100 \cMpch$
with each DM particle having a mass of $4.9\times 10^8 \Msunh$.  \par
Initial conditions for the particle's position and velocity were
obtained from glass-like initial conditions using \cmbfast\ (version
4.1; \citet{seljak96}) and the Zeldovich approximation was used to
linearly evolve the particles down to redshift z = 127. A flat
$\Lambda$CDM universe is assumed with the set of cosmological
parameters $\Omega_m = 0.238$, $\Omega_b = 0.0418$, $\Omega_\Lambda =
0.762$, $\sigma_8 = 0.74$, $n_s=0.951$ and $h=0.73$
\citep{spergel07}. Snapshots were written out at $35$ equally spaced
redshifts between $z = 12$ and $z = 6$.  Halos were identified using
the Friends-of-Friends algorithm (\citealp{davis85}), with linking
length $b = 0.2$. The smallest mass haloes we can resolve are a few
times $10^{10} \Msunh$. The density field was smoothed on the mesh
with a Gaussian kernel with standard deviation $\sigma_G = 2 \times
100 / 512$ comoving ${\rm Mpc}/h$ (2 grid cells).

\subsection{1-D radiative transfer (RT) code.}

The implementation of the 1-D radiative transfer code is modular and
hence straightforward to include different spectra corresponding to
different ionizing sources. Our 1-D code can handle X-ray photons, the
secondary ionization and heating it causes, and as a result perform
well both for high (quasars/miniqsos) and low energy (stars)
photons. Further details of the code can be found in \citet{thomas08}.

The parameter space simulated by the 1-D radiative transfer scheme
\citep{thomas08} include; 1) different SEDs (stars or miniqsos with
different power-law indices), 2) duration of evolution (depending on
the life-time of the source), 3) redshifts at which these sources turn
on and 4) clumping factor or over-density around a given
source. Information on the ionization and temperature profiles from
these simulations were cataloged to be used later as in
\citet{thomas09}. Density profiles around sources are assumed
  to be homogeneous although the code could be initialized using any
  given profile.

\subsection{BEARS algorithm: overview}
\textsc{BEARS} is a fast algorithm that simulates the underlying
cosmological 21-cm signal from the EoR. It is implemented by using the
Nbody/SPH simulation in conjunction with a 1-D radiative transfer
code; both discussed in the previous sections. The basic steps of the
algorithm are as follows: first, a catalogue of 1-D ionization profiles
of all atomic hydrogen and helium species and the temperature profile
that surround the source is calculated for different types of ionizing
sources with varying masses, luminosities at different
redshifts. Subsequently, photon rates emanating from dark matter
haloes, identified in the N-body simulation, are calculated semi -
analytically. Finally, given the spectrum, luminosity and the density
around the source, a spherical ionization bubble is embedded around
the source, whose radial profile is selected from the catalogue as
generated above. In case of overlap between ionization bubbles, the
excess ionized atoms in the overlap area is redistributed at the edge
of the bubbles. For more details see Thomas et al. (2009).  The outputs
are data cubes (2-D slices along the frequency/redshift
direction) of density ($\delta$), radial velocity ($v_r$) and hydrogen
and helium fractions ($x_\mathrm{HI; HII; HeI; HeII \& HeIII}$). Each
data cube consists of a large number of slices each representing a
certain redshift between 6 and 11.5.  Because these slices are
produced to simulate a mock dataset for radio-interferometric
experiments, they are uniformly spaced in frequency (therefore, not
uniform in redshift). Thus, the frequency resolution of the instrument
dictates the scales over which structures in the Universe are
averaged/smoothed along the redshift direction. The relation between
frequency and redshift space z is given by:$ z = \frac{\nu_{21}}{\nu}
-1$, where $\nu_{21} = 1420$ MHz is the rest frequency that
corresponds to the 21-cm line.

\section{Including IGM heating in \textsc{bears}}
\label{sec:bearheat}

In \cite{thomas09} the \textsc{bears} algorithm was introduced; a
special-purpose 3-D radiative transfer scheme used to simulate
cosmological EoR signals. In that paper we detailed the philosophy
behind the code and its implementation to incorporate ionization due
to a variety of sources. In this section, we extend the features of
\textsc{bears} to include heating (on which the brightness temperature
is sensitively dependent) due to hard photons within the same
framework.

The algorithm to implement heating begins in a manner similar to that
for ionization. The $\mathrm{T_k}$-profile from the 1-D radiative
transfer code is used to embed a spherically symmetric
``$\mathrm{T_k}$ bubble'' at the locations of the centre-of-mass of
dark matter haloes.  The luminosity of the source is a function of the
halo mass. The problem embedding a temperature profile in the
simulation box is that, the radius of the bubble is large ($>5$~Mpc)
and results in extensive overlap, even at high redshifts; a problem
encountered for ionization bubbles only towards the end of
reionization. In a quasar dominated part of
the Universe this is particularly severe because of its large
sphere-of-influence in heating the IGM.

\subsection{Treating overlaps: Energy conservation}

Overlap in the spheres of ionized regions was treated by
redistributing the excess ionizing photons in the overlapped region
uniformly around the overlapping spheres \citep{thomas09}. Treating
overlapped zones directly in terms of temperatures makes it difficult
to come up with any ``conserved quantity''. To elaborate, consider a
pre-ionized zone in the simulation box close to the radiating
source. Although at close proximity to the source, this region is not
heated efficiently because of the absence of neutral hydrogen to
capture the photons. On the other hand, the same source can dump more
energy into an initially neutral zone that maybe far away from
it. Thus, embedding the $\mathrm{T_k}$-profile directly from the
catalog (which assumes uniform IGM density) is not appropriate. 

Here  we adopt a different approach where we estimate the 
total energy deposited in every region of the simulation 
box and invoke the conservation of energy deposited by the sources in the 
overlapped regions. We integrate the contribution to the
energy budget, at a given location, from all sources whose
sphere-of-influence extents to that location. This total energy is
then redistributed equally to all the contributing sources, i.e., the
energy output of each source is modified to account for the
overlapping.

To illustrate, consider $\cal{N}$ sources that overlap at a particular
location $\vec{r}$ and the total energy estimated at that location to
be $E_{tot} = \sum_{i=1}^N E_i$, where $E_i$ is the energy deposited
by the $i^{th}$ source at location $\vec{r}$. The fraction of
``excess'' energy attributed to each of the $\cal{N}$ sources is
$\delta E = E_{tot}/\cal{N}$. This energy $\delta E$ is then added to
the total energy from the source to estimate a new normalization
constant for the SED of the source. For example in case of a
power-law-type source, the SED includes a normalization term that is
determined by the total energy of the source, $E_{total}$. However, in
the overlap case the total energy of the source is replaced by
$E_{total} + \delta E$. Now a ``new'' bubble of kinetic temperature is
embedded whose profile depends on a source whose output energy has
gotten a $\delta E$ boost. In this manner $\mathrm{T_k}(\vec{r})$ is
estimated at every location in the simulation box. From this point on
we need to estimate the coefficients that couple $\mathrm{T_k}$ to
$\mathrm{T_s}$ that will eventually lead to the estimation of $\delta
\mathrm{T_b}$.  It is worthwhile also to note that the results are
robust to variation in the exact algorithm used to treat the
overlapped region.  For example, if we choose to sum the energies
deposited directly in a given region or choose max($E_1,
E_2,...,E_n$), where $E_1...E_n$ are energies deposited by sources
1...n respectively, the final results are very similar.

This is obviously an approximation which assumes that the efficiency
of the photons in heating the IGM does not depend on their frequency.
For the case of power-law sources where most of the energy comes from
hard photons this is probably a reasonable approximation.  However, in
the case of blackbody sources this is not a very good assumption
because evidently the heating is done by a very small fraction of
photons. But in these cases the overlap problem is not as severe as
for the power-law sources since the extent of the heating is much more
limited.  In summary, despite the approximate nature of this approach, 
it provides an efficient and reasonable resolution for the issue of
the overlapping of temperature bubbles.  We note that self consistent
treatment of this problem requires taking into account the
modification of each of sources' SED in a different manner depending
on the amplitude of its SED. Obviously this is computationally 
very expensive.

\subsection{Calculating $\delta \mathrm{T_b}$ in the volume}

Spin temperature can be thought of as a short hand notation to
represent the level population of the hyperfine states of the ground
level of a hydrogen atom. Depending on the physical processes and
background radiation that dominate a medium, $\mathrm{T_s}$ is either
coupled to the background CMB temperature or to the kinetic
temperature of the hydrogen gas in the medium.  Formally
\cite{field58} derived $\mathrm{\mathrm{T_s}}$ as a weighted sum of
$\mathrm{\mathrm{T_k}}$ and $\mathrm{\mathrm{T_{CMB}}}$, as;

\begin{equation}
\mathrm{T_s^{-1}} = \frac{\mathrm{T_{CMB}^{-1}} (z)+y_\alpha \mathrm{T_k^{-1}} + y_{coll}\mathrm{T_k^{-1}}}{1+y_\alpha+y_{coll}},
\label{eq:tspin}
\end{equation}
where $y_\alpha$ and $y_{coll}$ are parameters that reflect the
coupling of $\mathrm{T_s}$ to $\mathrm{T_k}$ via Ly$\alpha$ excitation
and collisions respectively. The efficiency of Ly$\alpha$-coupling
dominates over that of collisions, especially further away from the
source \citep{thomas08,chuzhoy06}. In our treatment of calculating
$\mathrm{\mathrm{T_s}}$, we estimate both $y_\alpha$ and $y_{coll}$.

The coefficient $y_{coll}$ is a function of $\mathrm{\mathrm{T_k}}$
and the ionized fraction of the medium, $\mathrm{x_{HII}}$. These
informations on ionization and heating are available from the
prescription outlined in the previous section. While simulating the
1-D profiles we simultaneously calculate $y_\alpha $ as;

\begin{equation}
y_\alpha = \frac{16\pi^2T_\star e^2f_{12}J_o(|r|)}{27A_{10}T_e m_e c}.
\label{eq:yalpha}
\end{equation}
Here, $J_o(|r|)$ is the Ly$\alpha$ flux density at distance $r$ from
the source. For miniqsos with high energy photons, Ly$\alpha$ coupling
is mainly caused by collisional excitation due to secondary electrons
\citep{chuzhoy06}. This process is accounted for by the following
integral,
\begin{equation}
J_o(|r|) = \frac{c}{4 \pi H(z) \nu_\alpha} n_{HI}(r) 
         \int\limits_{50eV}^{2KeV} \phi_\alpha (E)\sigma(E) N(E;r,t) dE, 
\label{eq:lyalphaflux}
\end{equation}
where $N(E;r;t)$ is the number of photons with energy 'E' at radius
'r', and time 't' per unit area \citep{zaroubi05}. This number is
obtained directly from the 1-D radiative transfer by taking into
account the the absorption due to the optical depth along the
line-of-sight to the source at a distance 'r'. The 1-D profiles
generated are catalouged also as a function of time. Therefore the
profiles that are embedded within the simulation are chosen to obey
causality \footnote{Of course the catalouge has been generated with a
  a particular time resolution and can be made finer for more accurate
  results}. The ionization cross-section of neutral hydrogen in the
ground state is given by $\sigma(E),$ $\mathrm{f_{12}} = 0.416$ is the
oscillator strength of the Ly$\alpha$ transition, $e$ \& $m_e$ are the
electron's charge and mass, respectively, $\phi_\alpha$ is the
fraction of the absorbed photon energy that goes into excitation
\citep{shull85,dijkstra04a}. Note here that the fitting
  formula is a function of both ionization fraction and energy as
  given in the appendix of \citet{dijkstra04a}.

For the case of stars, the dominant source of Ly$\alpha$ flux results
from the redshifting of the source spectrum blueward of Ly$\alpha$
into the resonant line at different distances from the
source. Frequency '$\nu$' at the redshift of emission (or the source)
'$z$' is redshifted into $\nu_{Ly\alpha}$ at redshift $z(|r|)$ such
that;
\begin{equation}
\nu = \nu_{Ly\alpha} \frac{1+z}{1+z(|r|)}.
\end{equation}
{ Thus, at every radius $r$ away from the source the Ly$\alpha$ flux
  $J_o(|r|) \propto N(E';r;t)/r^2$. Where $E' = 10.2 ~
  \frac{1+z}{1+z(|r|)}$ [eV].  }

Now, instead of embedding a sphere of the $\mathrm{T_s}$, as in the
case of $\mathrm{T_k}$, we embed instead the a bubble of Ly$\alpha$
flux, $J_o(|r|)$, as estimated from Eq.~\ref{eq:lyalphaflux}. Since
$J_o(|r|)$ is basically the number of Ly$\alpha$ photons at a given
location, the overlap of two ``$\mathrm{J_o}$ bubbles'' implies that the
photons and hence the $\mathrm{J_o}$ has to be added, i.e., at a given
spatial (pixel) location, $\vec{x}$, $\vec{y}$, $\vec{z}$ and time
$t$, the total Ly$\alpha$ flux is given by;

\begin{equation}
J_{o}^{tot}(\vec{x},\vec{y},\vec{z},t) = \sum_{i=0}^N
J_{o}^{i}(\vec{x},\vec{y},\vec{z},t),
\label{eq:jtot}
\end{equation}
where $J_{o}^{i}(\vec{x},\vec{y},\vec{z},t)$ is the Ly$\alpha$ flux
contributed by $i^{th}$ source at the pixel location in the box,
$\vec{x}$, $\vec{y}$, $\vec{z}$ and time $t$. Equipped with the
quantities $\mathrm{\mathrm{T_k}(\vec{r})}$, and $J_o(\vec{r})$, we can
calculate $y_\alpha$ as in Eq.~\ref{eq:yalpha}, and subsequently the
spin temperature through Eq.~\ref{eq:tspin}. Now all terms required
for the calculation of $\delta \mathrm{T_b}$, as in Eq.~\ref{eq:dtb}, are
obtained.

\begin{eqnarray}
\delta \mathrm{T_b} (\vec{r}) =
\left(20~\mathrm{mK}\right) (1+\delta(\vec{r}))\left(\frac{\mathrm{x_{HI}}(\vec{r})}{h}\right)\left(1-\frac{\mathrm{T_{CMB}}}{\mathrm{T_s}(\vec{r})}\right)
\nonumber & & \\ \times \left[ \frac{H(z)/{1+z}}{dv_{||}/dr_{||}}\right] \left(\frac{\Omega_b h^2}{0.0223}\right)
\left[\left(\frac{1 +
z}{10}\right)\left(\frac{0.24}{\Omega_m}\right)\right]^{1/2} & &,
\label{eq:dtb}
\end{eqnarray}

\section{Example applications of {\textsc{bears}}}
\label{sec:applications}

In this section we apply \textsc{bears} with its extended feature of
including heating of the IGM, to three different scenarios of
reionization. The models described in this section are not
``template'' reionization scenarios by any measure nor is any
particular model favoured w.r.t the other. In fact, these models may
be far from reality and only serve as examples of the potential of
\textsc{bears} to provide a reasonable estimate of the 21-cm brightness
temperature for widely different scenarios of reionization.

\subsection{Heating due to stars}

The first scenario explored using \textsc{bears} is the case in which
the sources of reionization are \emph{only} stars. In this section we
describe the model used to describe the stellar component and the
prescription adopted to embed these stars into haloes of dark matter
identified in an N-body simulation.

\subsubsection{Modeling stellar radiation in \textsc{bears}}
\label{sec:stellar}
Most stars, to first order, behave as blackbodies at a given
temperature, although the detailed features in the SED depends on more
complex physical processes, age, metallicity and mass of the
star. This blackbody nature imprints characteristic signatures on the
IGM heating and ionization patterns \citep{thomas08}.
\cite{schaerer02} showed that the temperature of the star only weakly
depends on its mass. The blackbody temperature of the stars in our
simulation was thus fixed at $5 \times 10^4~\mathrm{K}$ to perform the
radiative transfer. We sample the parametre space of redshifts (12 to
6), density profiles around the source \footnote{We can incorporate
  any given density profile like the homogeneous, isothermal or NFW
  profiles. For this paper however we ran all the simulations using a
  homogeneous background. Overdensities around sources can be corrected
  for by using the prescribtion in \cite{thomas09}.} and masses (10 to
1000 $\mathrm{M_\odot}$). The total luminosity is normalized depending
on the mass of the star according to Table.~3 in \citet{schaerer02}.

For blackbody with a temperature around $10^5~\mathrm{K}$, similar to
our adopted value, the spectrum peaks between $\approx 20~eV $ to
$\approx 24~eV$ (Wien's displacement law). Thus, from the form of the
blackbody spectrum we can apriori expect, in the case of stars, that
the ionization induced by these objects will be high, but the heating
they cause will not be substantial given the exponential cutoff of the
radiation towards higher frequencies.

Now, following the prescription in \cite{thomas09} we associate
stellar spectra with dark matter halos using the following
procedure. The global star formation rate density
$\dot{\rho_\star}(z)$ as a function of redshift was calculated using
the empirical fit;

\begin{equation}
\dot{\rho_\star}(z) = \dot{\rho_m} \frac{\beta\exp\left[\alpha(z - z_m)\right]}{\beta - \alpha + \alpha\exp\left[\beta(z - z_m)\right]}~~[{\rm M}_\odot~{\rm yr}^{-1}~{\rm Mpc}^{-3}],
\label{eq:starformrate} 
\end{equation}
where $\alpha= 3/5$, $\beta=14/15$, $z_m=5.4$ marks a break redshift,
and $\dot\rho_m= 0.15 [{\rm M}_\odot~{\rm yr}^{-1}{\rm Mpc}^{-3}]$
fixes the overall normalisation \citep{springel03}. Now, if $\delta t$
is the time-interval between two outputs in years, the total mass
density of stars formed is
\begin{equation}
\rho_\star (z)\approx \dot\rho_\star(z)\delta t ~~[{\rm M}_\odot~{\rm Mpc}^{-3}].
\end{equation} 
Notice that this approximation is valid only if the typical lifetime
of the star is much smaller than $\delta t$, which indeed is the case
in our model because the stellar mass we input (100 $\mathrm{M_\odot}$) has a
lifetime of about a few Myrs \citep{schaerer02}, compared to a typical
time difference of few tens of Myrs between simulation snapshots.

Therefore, the total mass in stars within the simulation box is $
M_\star(box) \approx L_{box}^3 \rho_\star ~~[{\rm M}_\odot]$. This
mass in stars is then distributed among the halos weighted by
their mass;
\begin{equation}
m_\star(halo) = \frac{m_\mathrm{halo}}{M_\mathrm{halo}(tot)} M_\star(box),
\end{equation}
where $m_\star(halo)$ is the mass in stars in the dark matter ``halo'',
$m_\mathrm{halo}$ the mass of the halo and $M_\mathrm{halo}(tot)$ the
total mass in halos within the simulation box.

We then assume that all of the mass in stars is distributed in stars
of 100 $\mathrm{M_\odot}$, which implies that the number of stars in the
halo is $N_{100} = 10^{-2} \times m_\star(halo)$. The luminosity of a
100 $\mathrm{M_\odot}$ star is obtained from Fig.~1 of \citet{schaerer02},
assuming zero metallicity. The luminosity of 100 $\mathrm{M_\odot}$ star thus
derived is in the range of $10^6$ --- $10^7~L_\odot$ and this value is
multiplied by $N_{100}$ to get the total luminosity emanating from the
``halo'' and the radiative transfer is done by normalizing the
blackbody spectrum to this value. The escape fractions of ionizing
photons from early galaxies is assumed to be 10\%.

\subsubsection{Results: Stellar sources}

\begin{figure}
\centering
\hspace{0cm}
\includegraphics[width=0.5\textwidth]{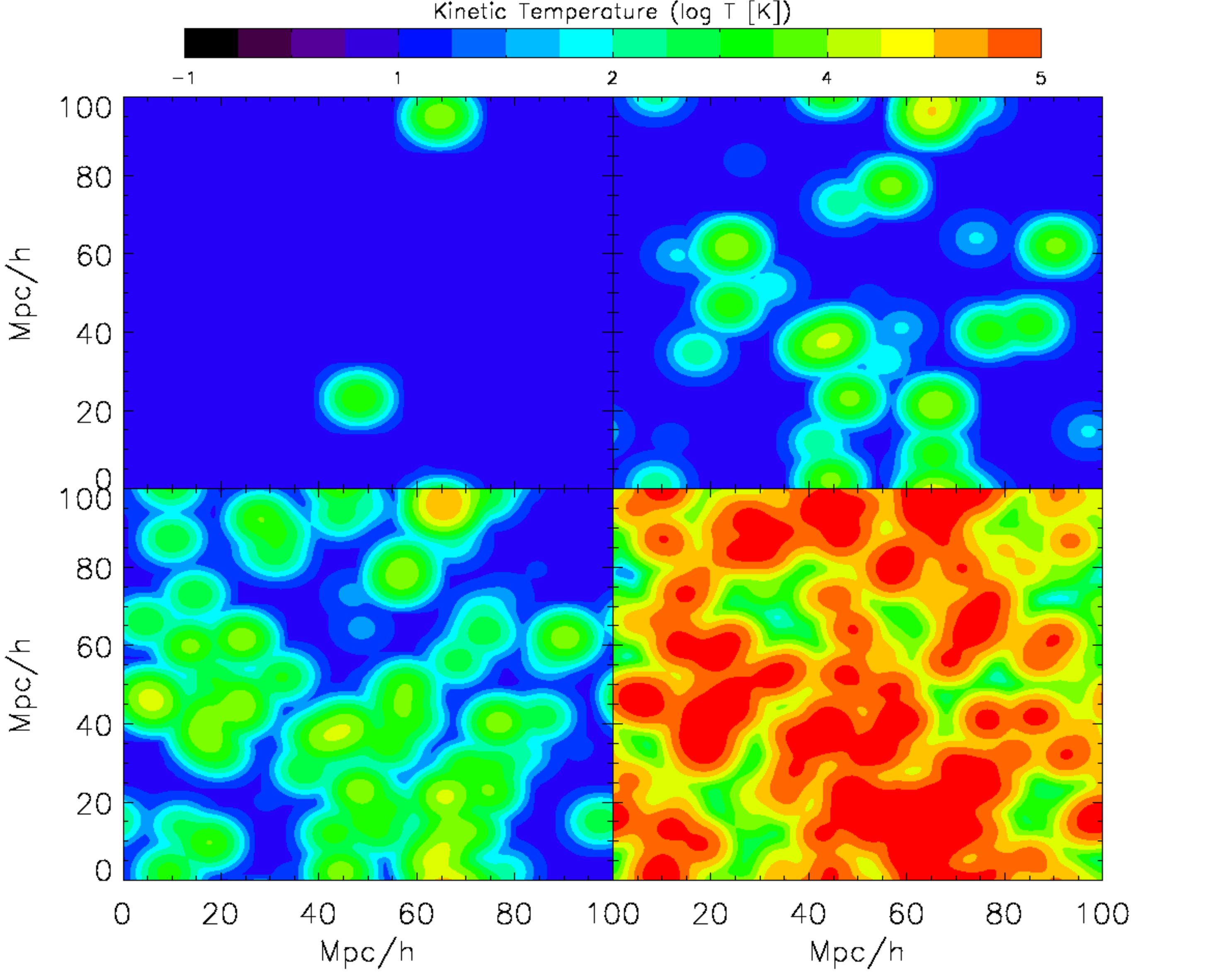}
\vspace{-.5cm}
\caption{\emph{Kinetic Temperature for stellar sources}: Slices of
  $\mathrm{T_k}$ corresponding to redshifts 10, 8, 7 and 6 are plotted
  (\emph{panels left to right and top to bottom, respectively}).
  Temperature are indicated in log T [K] on the color bar above. The
  extent of heating, both in amplitude (maximum around $\approx 10^5$
  K towards the center) and spatial extent ($<$ 100 kpc), is extremely
  small. Basically, only the central part (ionized region) is at a
  high temperature and falls sharply during the transition into the
  neutral IGM.}
\label{fig:starheat}
\end{figure}

\begin{figure}
\centering
\hspace{0cm}
\includegraphics[width=0.5\textwidth]{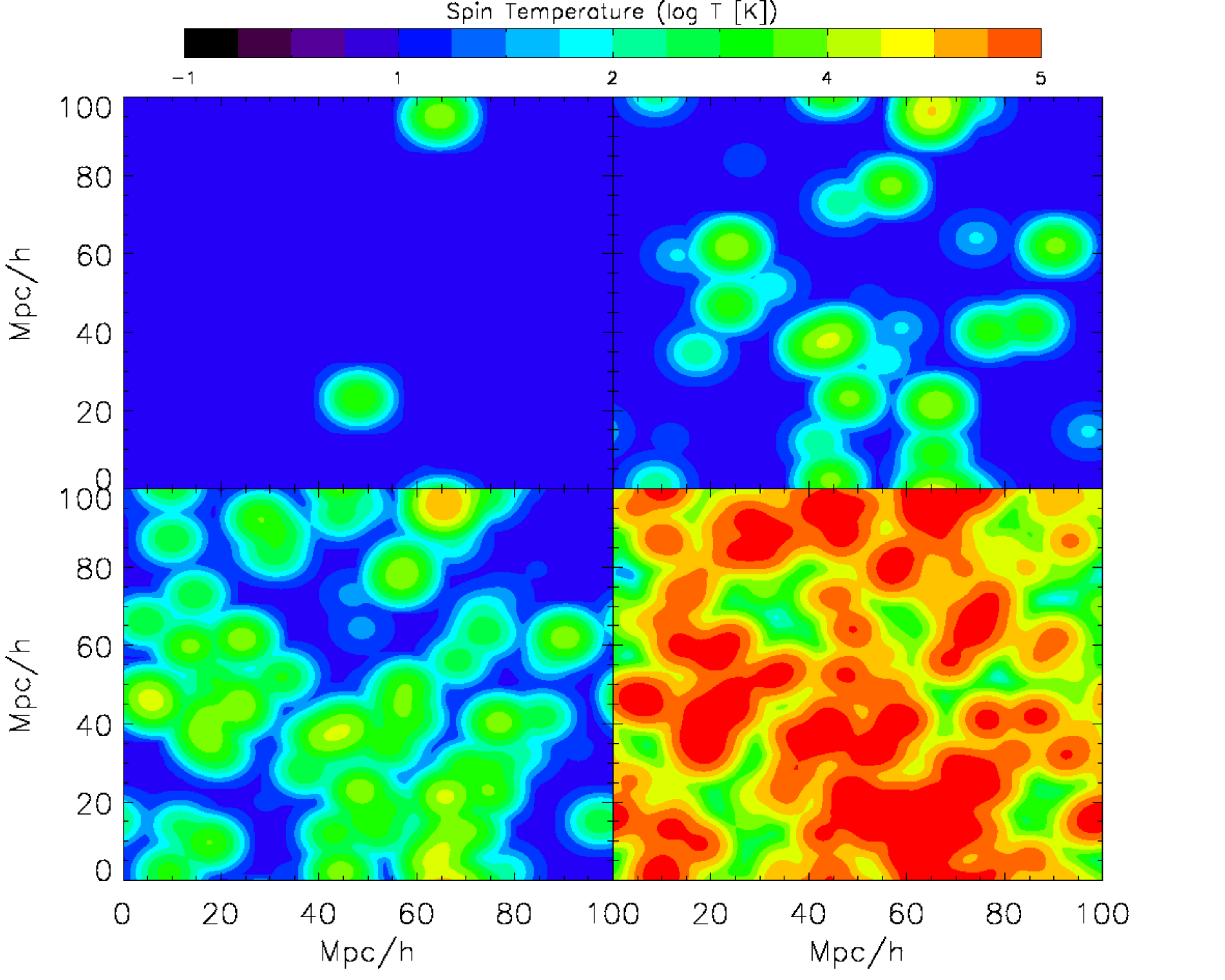}
\vspace{-.5cm}
\caption{\emph{Spin Temperature for stellar sources}: Slices
  correspond to that in Fig.~\ref{fig:starheat}. Collisional coupling
  is efficient to couple the $\mathrm{T_k}$ to $\mathrm{T_s}$ at the
  centre but the figures also indicate that there is sufficient
  $\mathrm{J_o}$ to couple $\mathrm{T_k}$ to $\mathrm{T_s}$ well
  beyond the ionized region, where $\mathrm{T_k} < \mathrm{T_{CMB}}$.
  The color bar indicates temperature on a log T [K] scale.}
\label{fig:starspin}
\end{figure}

\begin{figure}
\centering
\hspace{0cm}
\includegraphics[width=0.5\textwidth]{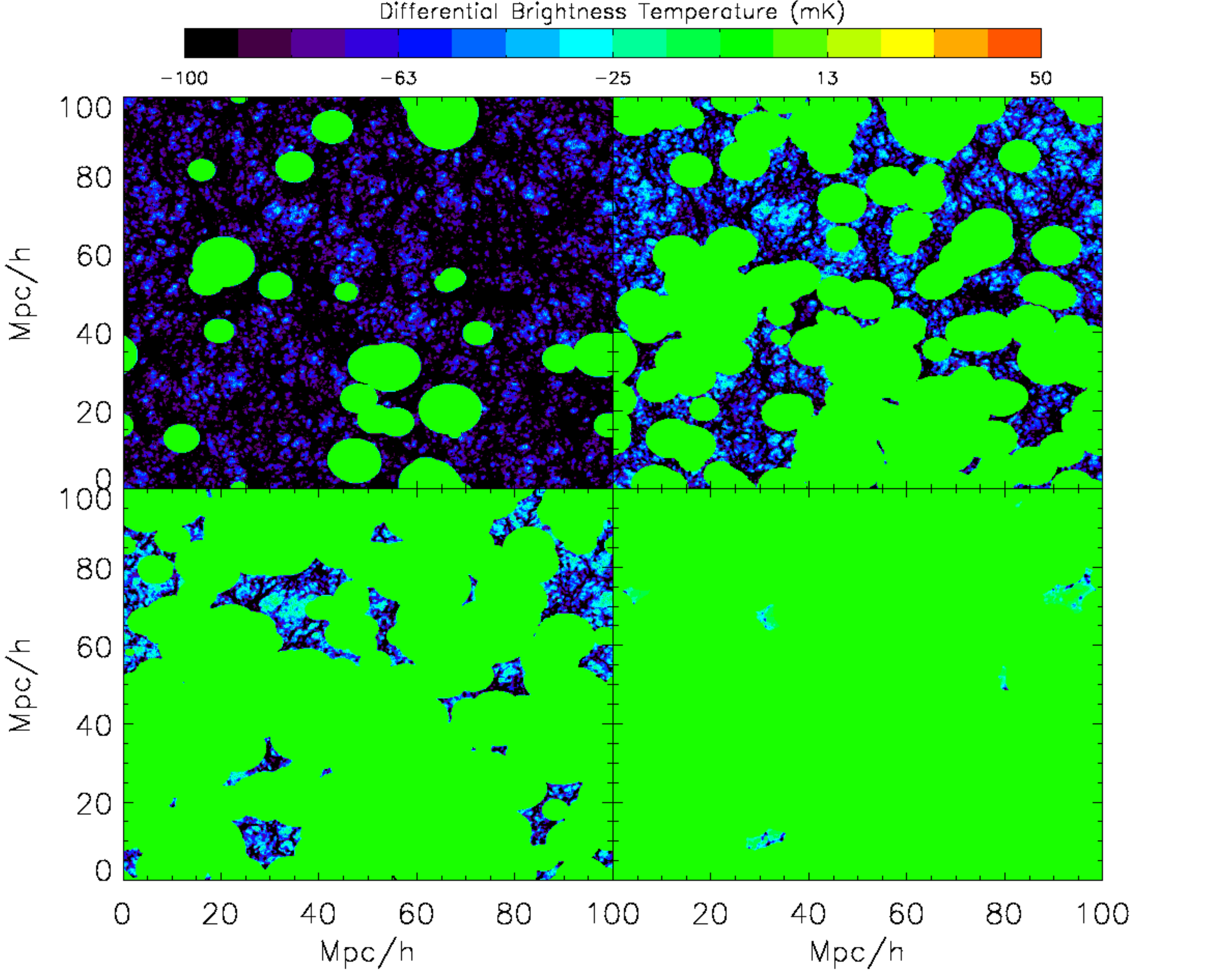}
\vspace{-.5cm}
\caption{\emph{Differential Brightness Temperature for stellar
    sources}: Slices corresponds to that in
  Fig.~\ref{fig:starheat}. The temperature is now contoured on a
  linear scale with values indicated on the colorbar. In these slices,
  $\delta \mathrm{T_b}$, starting from zero within ionized regions
  attains large negative value outside because stars are not efficient
  at heating the IGM beyond their ionized bubble and hence cannot
  render the IGM visible in emission. Early on, (top-left) at a
  redshift of 10, there are only a few sources and the heating induced
  colliosional coupling or the $\mathrm{J_o}$ is not high enough to
  render the Universe visible, but the photons blueward of Ly$\alpha$
  can reach far outside the ionized region where they are injected
  into Ly$\alpha$ frequency causing substantial coupling to the
  kinetic temperature which is much below the CMB. And, at later
  times, the ionized bubbles overlap significantly driving the
  brightness temperature to zero.}
\label{fig:starbright}
\end{figure}

Results of the evolution of the $\mathrm{T_k}$, $\mathrm{T_s}$ and
$\delta \mathrm{T_b}$ of the IGM, when the sources of reionization
comprise only of stars is shown in Figs \ref{fig:starheat},
\ref{fig:starspin} and \ref{fig:starbright}, respectively.

Figure \ref{fig:starheat} shows the kinetic temperature for stellar
sources at redshifts 10, 8, 7 and 6 at which their ionized fractions
are 0.12, 0.5, 0.83, and 0.98 respectively. The blackbody type stellar
spectra do not have sufficient high energy photons to heat the IGM
substantially far away from the source.  Thus we see compact regions
(bubbles) of high temperature in the immediate vicinity of the
source. In the inner parts, where the ionized fraction is high
($\mathrm{x_{HII}}>0.95$), the temperatures are of the order of
$\approx 10^5$ K and drops sharply in the transition zone to neutral
IGM. The ionized region is restricted to about 100 kpc in physical
coordinates around a halo \citep{thomas08}.

Figure.~\ref{fig:starspin} shows $\mathrm{T_s}$ at four different
redshifts. Collisional coupling ($y_{coll}$) is important close to the
radiating source ($< 200~kpc$). The primary source of coupling
  between $\mathrm{\mathrm{T_s}}$ and $\mathrm{\mathrm{T_k}}$ is the
  redshifting of photons blueward of the Ly$\alpha$ line into the
  Ly$\alpha$ frequency. It has to be emphasized here that the
Ly$\alpha$ photons are produced \emph{only} due to the source and
secondary processes. Thus we see that stellar sources are efficient in
building up a significant background of $\mathrm{J_o}$ far away from
the source which is sufficient to couple $\mathrm{\mathrm{T_s}}$ to
$\mathrm{\mathrm{T_k}}$ during reionization \citep{ciardi03}. As an
example, in Fig.~\ref{fig:histcompinf} we show the reionization
history ($\delta \mathrm{T_b}$) assuming a high background
$\mathrm{J_o}$, in other words, assuming perfect
$\mathrm{T_s}$---$\mathrm{T_k}$ coupling.

In Fig.~\ref{fig:starbright} $\delta \mathrm{T_b}$ corresponding to
the same redshifts as in the previous figures is shown. In this
particular model $\delta \mathrm{T_b}$ values are not very high
($|\delta \mathrm{T_b}| < 10$mK). At early epochs, (top-left) at
$z\approx10$, there are only a few sources and the heating of the IGM
or the secondary $\mathrm{J_o}$ is not high enough to cause substantial
differential brightness temperatures. At later times, the ionized
bubbles overlap significantly driving the brightness temperature to
zero. Thus there is only a small portion of the Universe (in volume)
that has sufficient Ly$\alpha$ coupling and neutral hydrogen density
to cause large differential brightness temperatures. This behaviour is
specific to the model assumed and will be significantly different if
the parameters such as the star-formation rate, SED and escape
fractions are altered.

\subsection{Heating due to miniqsos}
\label{subsec:quasarheat}

In this section we will consider heating due to high energy
X-ray photons emanating from power-law type sources (eg.,
miniqsos). Because of their large mean free path ($\propto E^3$) it
has been difficult to incorporate the effect of heating by power-law
sources self-consistently in a 3-D radiative transfer simulation.

\subsubsection{Modeling miniqsos in \textsc{bears}}

Observations reveal the energy spectrum of quasars-type sources
typically follow a power-law of the form $E^{-\alpha}$
\citep{vandenberk01,vignali03,laor97,elvis94}. Here we assume $\alpha
= 1$ and thus the SED of the miniqsos is given by:

\begin{equation}
I(E) = Ag \times E^{-\alpha} ~~10.4~\mathrm{eV} < E < 10^4~\mathrm{eV},
\label{eq:spectrum}
\end{equation}
with the normalization constant, 
\begin{equation}
Ag = \frac{E_{total}}{\int\limits_{E_{range}} I(E) \mathrm{dE}},
\label{eq:normalization}
\end{equation}
where $E_{total}$ is the total energy output of the miniqso within the
energy range, $E_{range}$. Any complex, multi-slope spectral templates
as in \cite{sazonov04} can be adopted. The number of plausible SEDs
that can be considered are numerous and serves as a reminder of the
extent of ``unexplored parameter space'' even while considering the
case of miniqsos alone, and argues for an extremely quick RT code like
\textsc{bears}.

The miniqsos are assumed to accrete at a constant fraction
$\epsilon$ (normally 10\%) of the Eddington rate. Therefore, the
luminosity is given by:

\begin{eqnarray}
L & = & \epsilon\, L_{\mathrm{edd}}(\mathrm{M})\\ & = & 1.38 \times 10^{37}
\left(\frac{\epsilon}{0.1}\right) \left(
\frac{\mathrm{M}}{\mathrm{M_\odot}}\right) \mathrm{erg~s^{-1}}.
\label{eq:eddington}
\end{eqnarray}

 The luminosity derived from the equation above is used to normalize
 the relation in Eq.~\ref{eq:spectrum} according to
 Eq.~\ref{eq:normalization}. The energy range considered is between
 10.4 $\mathrm{eV}$ and 10 $\mathrm{keV}$.  Simulations were carried
 out for a range of masses between $10^5$ and $10^9~\mathrm{M}_\odot$.
 To justify comparing the case of stars and miniqsos we argue that,
 although the number of photons at different energies is a function of
 the total luminosity and spectral index in the case of miniqsos, if
 we assume that all photons from the miniqso are at the hydrogen
 ionization threshold, then the number of ionizing photons obtained,
 for the mass range fixed above, is about
 $10^{50}~\mathrm{to}~10^{55}$; the same order of magnitude as the
 number of ionizing photons in the case of stars and it also matches
 the numbers being employed for simulations by various other authors
 like \cite{mellema06} and \cite{kuhlen05}.

The miniqsos are embedded into an N-body output following the
prescription in \cite{thomas09}. Dark-matter haloes are identified
using the friends-of-friends (FoF) algorithm and masses of black holes
assigned according to,

\begin{equation}
 \mathrm{M_{BH}}= 10^{-4}\times\frac{\Omega_b}{\Omega_m}M_{halo},
\label{eq:mbh}
\end{equation}
 where the factor $10^{-4}$ reflects the Magorrian relation
 \citep{magorrian98} between the halo mass ($\mathrm{M_{halo}}$) and
 black hole mass ($\mathrm{M_{BH}}$) multiplied with the radiative
 efficiency assumed to be 10\%, and the fraction
 $\frac{\Omega_b}{\Omega_m}$ gives the baryon ratio
 \citep{ferrarese02}.

\subsubsection{Results: miniqsos}
\label{subsec:plsrcs}
\begin{figure}
\centering
\hspace{0cm}
\includegraphics[width=0.5\textwidth]{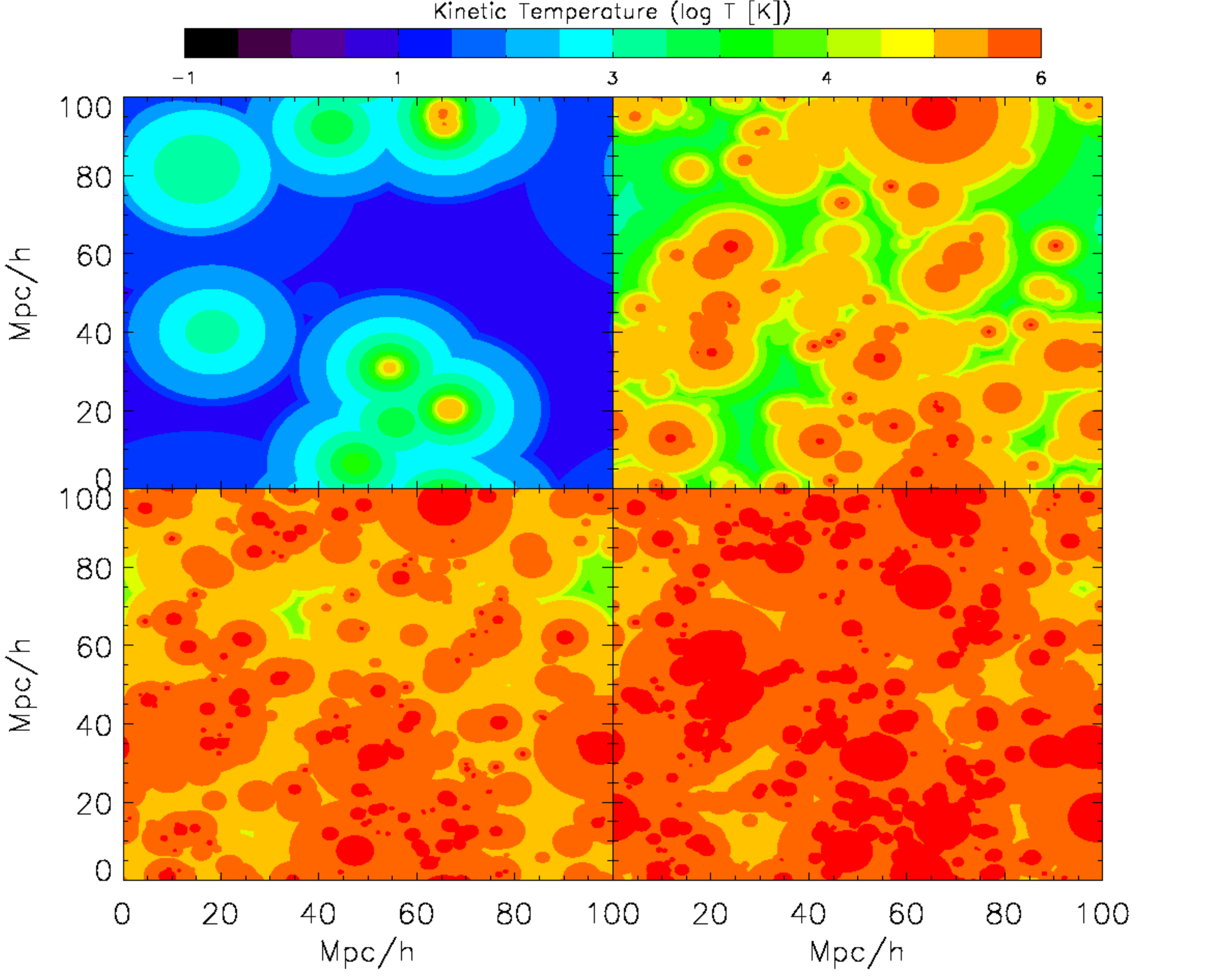}
\vspace{-.5cm}
\caption{\emph{Kinetic Temperature for miniqso sources}: Slices
  correspond to that in Fig.~\ref{fig:starheat} with temperatures
  indicated on a log T [K] scale. Here we see the striking difference
  between miniqsos and stars in the extent of heating. The amplitude
  is about $\approx 10^{6.5}$ K towards the center and spatial extent
  of the heating is about a few Mpcs. Already at redshift 10
  (top-left) the average temperature is of the order $10^3$ K.}
\label{fig:qsoheat}
\end{figure}

\begin{figure}
\centering
\hspace{0cm}
\includegraphics[width=0.5\textwidth]{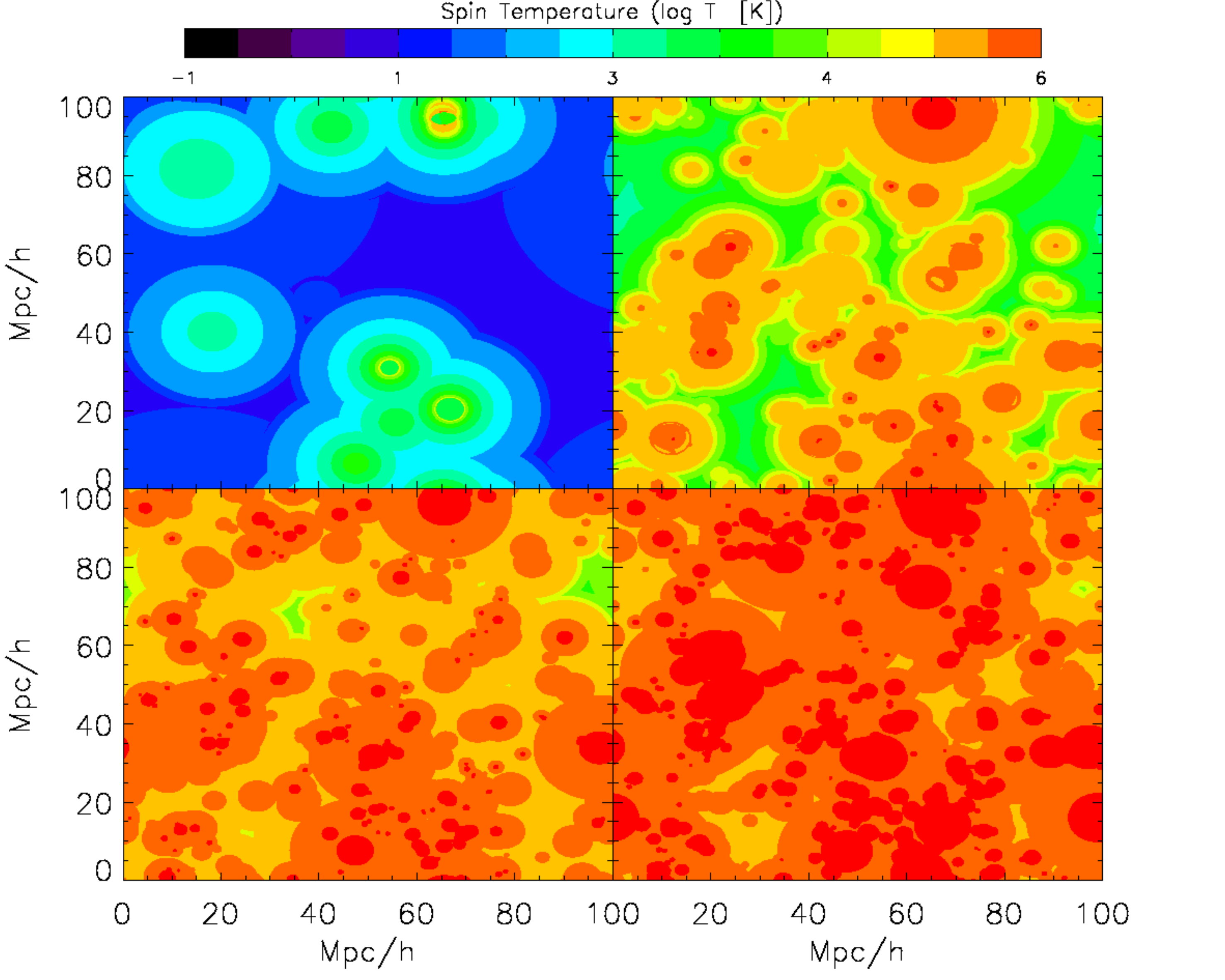}
\vspace{-.5cm}
\caption{\emph{Spin Temperature for miniqso sources}: Slices
  correspond to that in Fig.~\ref{fig:starheat} with the temperatures
  indicated on a log T [K] scale. Spin temperatures show an interesting
  ring-like behaviour (easily visible in the top-left panel).  The
  reason for this is discussed in the text.  }
\label{fig:qsospin}
\end{figure}

\begin{figure}
\centering
\hspace{0cm}
\includegraphics[width=0.5\textwidth]{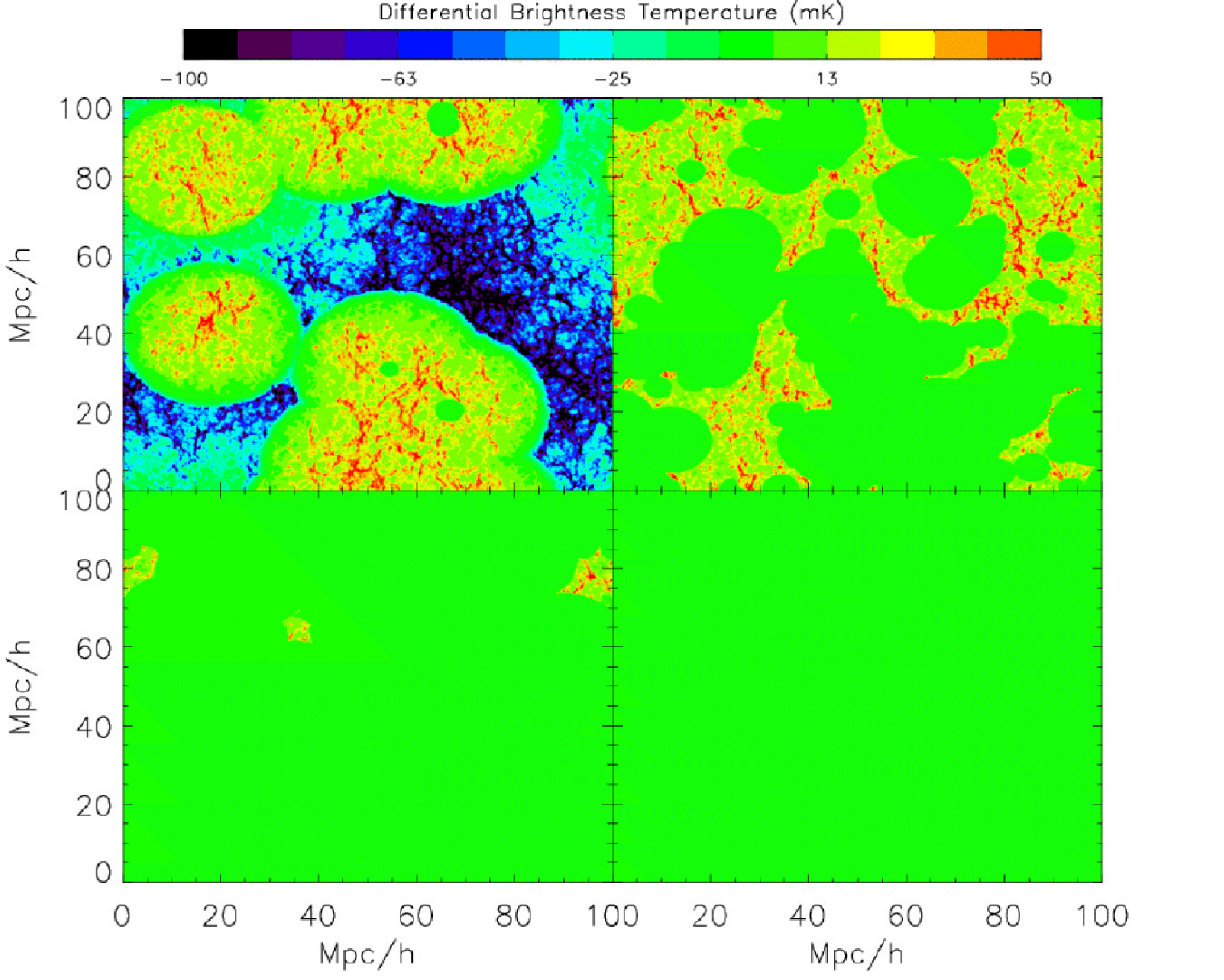}
\vspace{-.5cm}
\caption{\emph{Differential Brightness Temperature for miniqso
    sources}: Slices correspond to that in
  Fig.~\ref{fig:starheat}. Note here that the brightness temperature
  (mK) in plotted on a linear scale with values as indicated on the
  color bar. We see again from the top-left panel that although the
  number of sources in the field are few, they were efficient in both
  increasing $\mathrm{T_k}$ dramatically and providing sufficient
  $\mathrm{J_o}$ to ``light-up'' the Universe in 21-cm around them. The spheres
  in the top-left panel have $\delta \mathrm{T_b}$ of zero because
  they are highly ionized. }
\label{fig:qsobright}
\end{figure}

Contour plots of $\mathrm{T_k}$, $\mathrm{T_s}$ and and $\delta
\mathrm{T_b}$ for the case of \emph{only} miniqsos being the sources
of reionization, at four different redshifts, are plotted in
Figs.~\ref{fig:qsoheat}, \ref{fig:qsospin} and \ref{fig:qsobright},
respectively. The high energies of the X-ray photons substantially
affects the photo-ionization and thermal energy balance of the IGM.

The marked difference between power-law type and stellar sources, in
the extent to which they heat the IGM, is evident in
Fig.~\ref{fig:qsoheat}. The temperatures are shown at redshifts 10, 8,
7 and 6 with ionized fractions 0.01, 0.54, 0.96, 0.999
respectively. In the inner parts of the ``$\mathrm{T_k}$ bubble'' the
amplitude of $\mathrm{T_k}$ is about $\approx 10^{6.5}$ K. The
secondary electrons created by high energy X-ray radiation thermalizes
to these high temperature values. Also contributing to extremely high
temperatures in the very inner parts of the bubble is Compton heating
\citep{thomas09}. The mean free path of X-rays are extremely large
resulting in a spatial extent of heating that reaches many Mpcs. Thus,
we see that already at redshift 10 (top-left) the average temperature
of the IGM is of the order of $10^3$ K and overlap of the temperature
bubbles occur much before that of the ionization bubble. This example
might be unrealistic, because the soft X-ray background constraint
\citep{dijkstra04a} is not strictly observed, but again this serves
just as an illustrative example.

Figure.~\ref{fig:qsospin} shows the evolution of $\mathrm{T_s}$ as a
function of redshift. Owing to a high secondary $\mathrm{J_o}$
produced due to X-ray photons, $\mathrm{T_s}$ is coupled to
$\mathrm{T_k}$ for the large part. The snapshots of $\mathrm{T_s}$
show a peculiar ring-like behaviour
(Refer. Fig.~\ref{fig:qsospin}). This is because towards the central
part of the bubble around the source, the IGM is highly ionized and
$\mathrm{J_o} \propto \mathrm{x_{HI}}$, obtains an extremely low
value. On the other hand $\mathrm{J_o}$ progressively gets lower away
from the source, regions ($>$5Mpc). Therefore only a relatively
narrow zone (few hundred kpcs) in between the ionizing front and
regions further out has $\mathrm{J_o}$  high enough to decouple
$\mathrm{T_s}$ from $\mathrm{T_{CMB}}$.

Figure.~\ref{fig:qsobright} shows $\delta \mathrm{T_b}$ for redshifts
10, 9, 8 and 7 in panels top-left to bottom-right, respectively. The
$\delta \mathrm{T_b}$ plotted in this figure are high enough to be
detectable by upcoming telescope's like LOFAR and MWA, whose
sensitivities should reach $\delta \mathrm{T_b} > 5 \mathrm{mK}$
\citep{panos09}. We see in the top-left panel of
this figure that even though the number of sources in the field is
small, they are efficient in both increasing $\mathrm{T_k}$
dramatically and providing sufficient $\mathrm{J_o}$ to ``light-up''
the Universe in the 21-cm differential brightness temperature. The
$\delta \mathrm{T_b}$ inside the spheres are zero because they are
highly ionized ($\mathrm{x_{HII}}>0.99$).

\subsection{Modeling co-evolution of stars \& miniqsos}
\label{subsec:starqso}

In the observable Universe we know that stars and quasars do
co-exist. Although, there are indications of the quasar number-density
peaking around redshift two \citep{nusser93} and declining thereof,
there are also measurements by \citet{fan06} of high-mass ($>10^8
\mathrm{M}_\odot$) quasars at $z > 6$. This allows us to
envisage a scenario of reionization in which stars and quasars (or
miniqsos) contributed to the ionization and heating of the IGM in a
combined fashion. As a final example-application of \textsc{bears}, we
present our model of a ``hybrid'' star-miniqso SED and examine its
results on the ionization and heating of the IGM.

\subsubsection{The ``Hybrid'' SED}

The uncertainty regarding the properties and distributions of objects
during the dark ages allows us to come up with strikingly different
ways of incorporating the co-existence of stars and miniqsos. For
example, consider that there are $\cal{N}$ different haloes within our
simulation box. One approach would be to suppose that each of these
haloes host both a miniqso and stars, whose masses/luminosity is
derived according to the previous two examples. The other approach
would be to place one or two massive miniqsos (almost quasars), still
conforming to the black hole mass density predictions of
\cite{volonteri08}, and the rest of the haloes in the simulation box
with populations of stars. For this example, given the minimum mass
halo our simulation can resolve, we have chosen to adopt the former
approach, i.e., a miniqso and stars in every halo.

In our approach, given the halo mass we embed them with stellar
sources as described in \S\ref{sec:stellar}. Then the black hole mass
is set as $10^{-4}$ times the stellar mass. Radiative transfer is
performed using this hybrid SED, i.e., a superposition of power-law
type miniqso SED and black body spectrum normalized according to the
criterion in the previous sections. The results of ionization and
heating are presented below.

\subsubsection{Results: Hybrid model}

\begin{figure}
\centering
\hspace{0cm}
\includegraphics[width=0.5\textwidth]{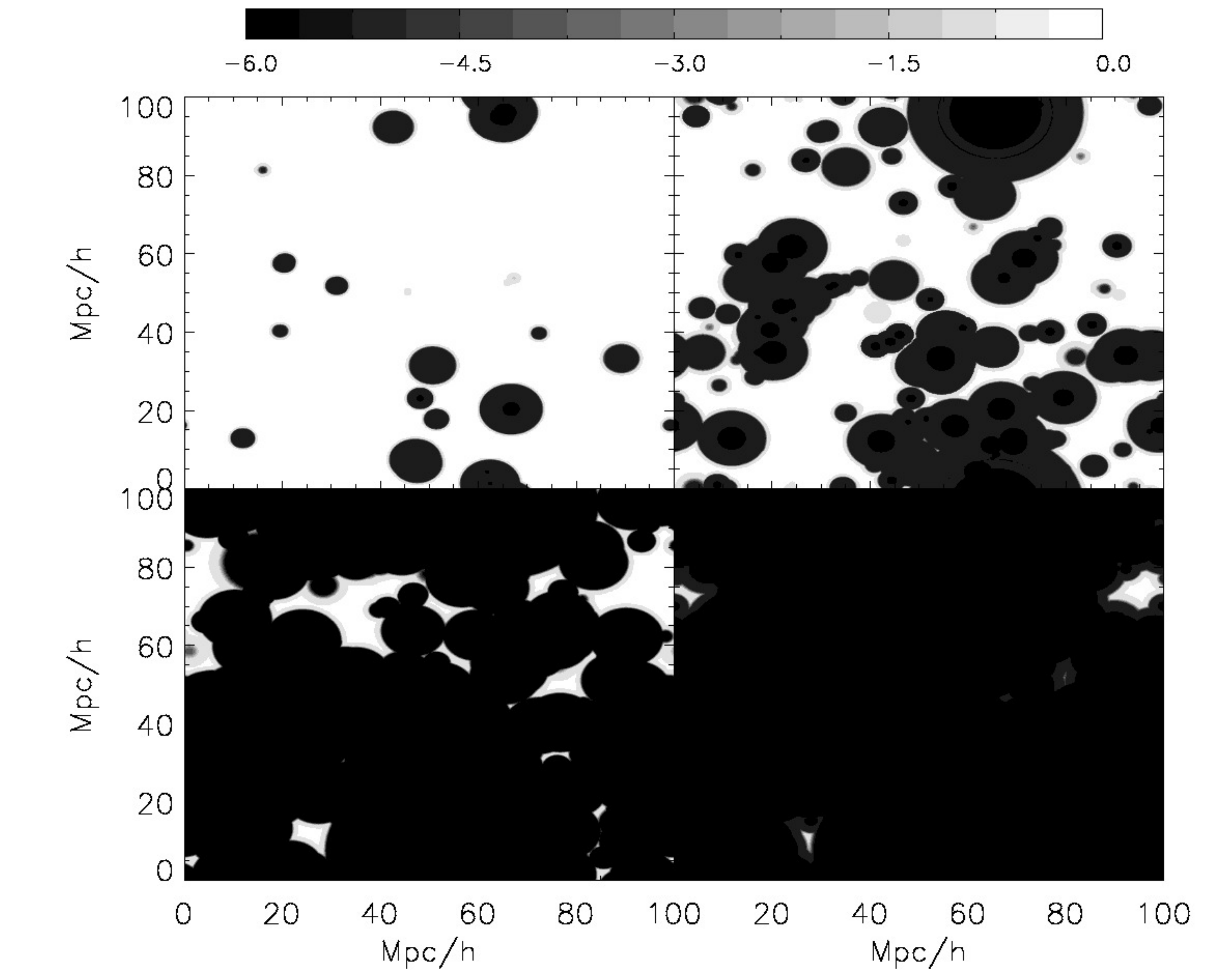}
\vspace{-.5cm}
\caption{\emph{Ionization due to the hybrid model}: Slices correspond
  to that in Fig.~\ref{fig:starheat}. Shown in these slices are the
  contour levels of neutral hydrogen in a logarithmic scale. We see
  here that for the hybrid model we have assumed the ionization
  proceeds quickly and the Universe is substantially ionized by
  redshift 6 (Refer Fig.~\ref{fig:histcomp}) for the reionization
  history. The extent of ionizations interpolates between the cases of
  black holes and stars.}
\label{fig:combiion}
\end{figure}

The hybrid model proposed does have large amounts of high-energy X-ray
photons from miniqsos that influence the temperature and ionization of
the IGM prior to full reionization. The X-rays photons have larger
mean free paths than their UV counterparts, and their secondary
electrons have significant influence on the thermal and ionization
balance. The UV photons on the other hand are efficient at keeping the
IGM in the vicinity of the halo highly ionized. For the redshifts 10,
8, 7 and 6 we plot the ionization due to the hybrid model
in Fig.~\ref{fig:combiion} and the ionized fraction at these redshifts are
0.01, 0.38, 0.86 and 0.995 respectively.

For this hybrid model the number of UV photons was not high enough to
ionize the Universe completely by redshift 6 (the ionized fraction
reaches 0.95). Full reionization thus could be achieved either by
increasing the mass of the black hole and/or by changing the ratio
between the black hole and stellar masses.
  
\begin{figure}
\centering
\hspace{0cm}
\includegraphics[width=0.5\textwidth]{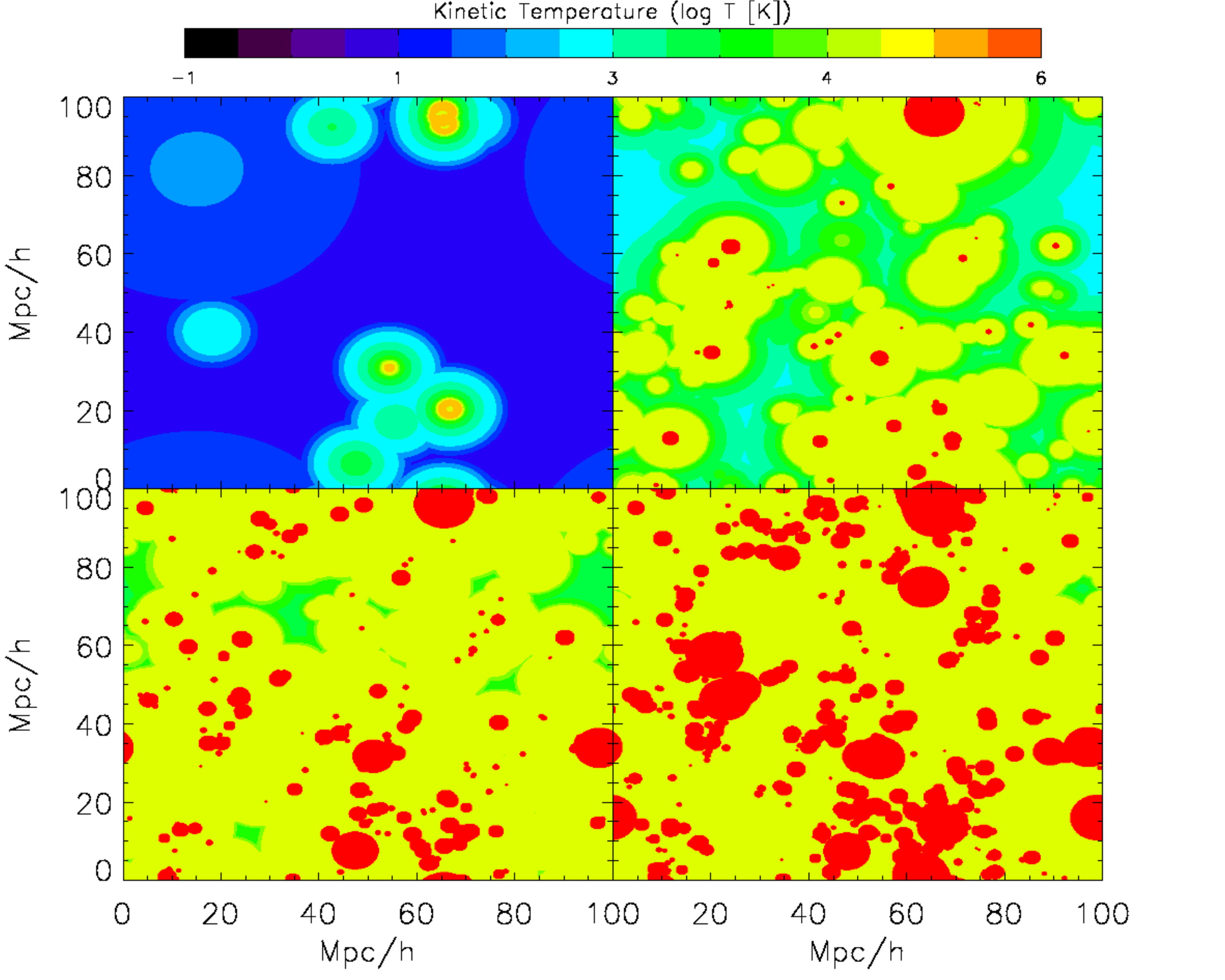}
\vspace{-.5cm}
\caption{\emph{Kinetic temperature for hybrid model}: Slices
  correspond to that in Fig.~\ref{fig:starheat}, with temperatures
  indicated on a log T [K] scale. The heating pattern is peculiar in that
  the temperature shows a much ``cuspier'' behaviour towards the
  center. The reason is that the stellar component of the source
  ionized the very central part and the X-ray radiation can thus heat
  the central part where to even higher temperature. On the other hand
  the large extent of the heating is attributed to the power-law
  component of the SED.}
\label{fig:combiheat}
\end{figure}

\begin{figure}
\centering
\hspace{0cm}
\includegraphics[width=0.5\textwidth]{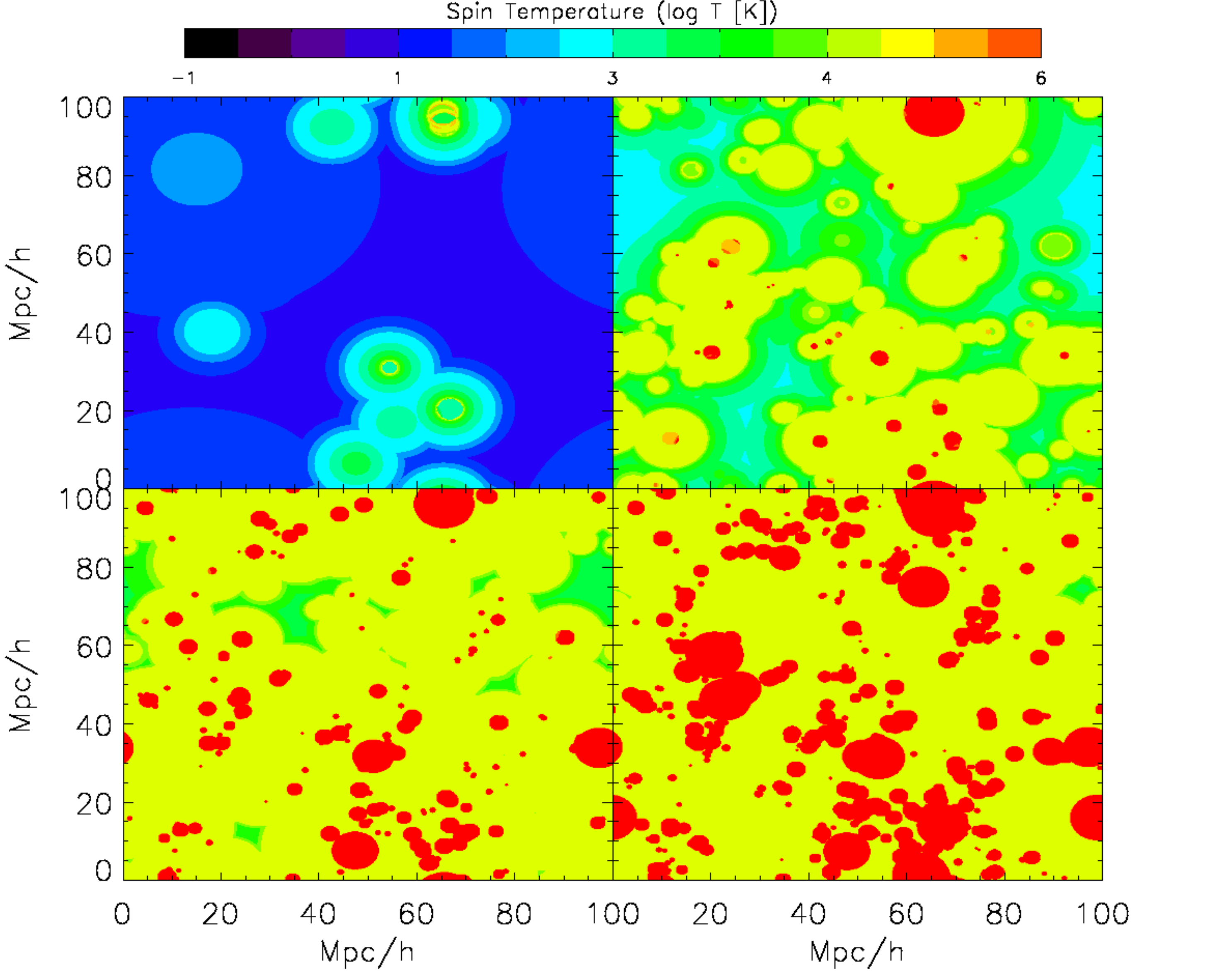}
\vspace{-.5cm}
\caption{\emph{Spin temperature for the hybrid model}: Slices
correspond to that in Fig.~\ref{fig:starheat}. Spin temperatures
for the case of the hybrid model again shows the ring like behaviour
as in the case of power-law sources. Temperatures are indicated
on a log T [K] scale.}
\label{fig:combispin}
\end{figure}

\begin{figure}
\centering
\hspace{0cm}
\includegraphics[width=0.5\textwidth]{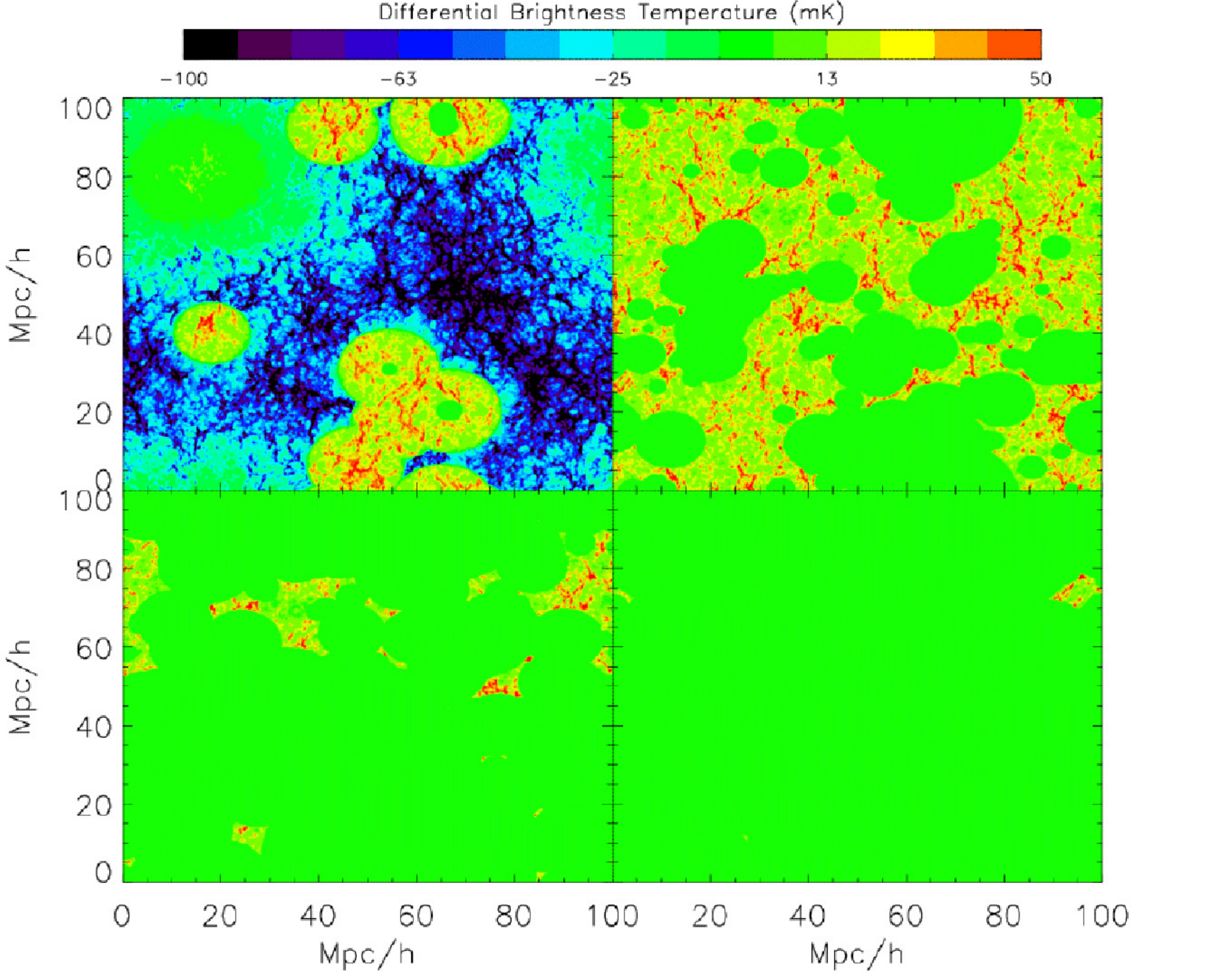}
\vspace{-.5cm}
\caption{\emph{Differential Brightness temperature for the hybrid
    model}: Slices correspond to that in Fig.~\ref{fig:starheat}. The
  temperatures (mK) are indicated on a linear scale in the color bar
  above. The large extent of heating and sufficient $\mathrm{J_o}$ ensures that
  the IGM well beyond the location of the source is rendered visible
  in 21-cm emission.  }
\label{fig:combibright}
\end{figure}

The hybrid model considered here yielded $\mathrm{T_k}$,
$\mathrm{T_s}$ and $\delta \mathrm{T_b}$ as in
Figs.~\ref{fig:combiheat}, Fig.~\ref{fig:combispin} and
Fig.~\ref{fig:combibright}, repectively.

The hybrid model produces heating patterns that interpolates between
the results for the previous two cases and thus is qualitatively
different (Fig.~\ref{fig:combiheat}).  The power-law components take
over the role of heating an extended region and the stellar component
maintains a very high temperature in the central parts.

The spin temperature (Fig.~\ref{fig:combispin}) again shows the
characteristic ring-like structure around the source. The amount of
secondary Ly$\alpha$ radiation produced due to the miniqsos results in
an efficient coupling of the spin temperature and the kinetic
temperature. 

The brightness temperature for the hybrid model is plotted in
Fig.~\ref{fig:combibright}. The secondary $\mathrm{J_o}$ produced
due to the power-law component efficiently couples the high ambient
$\mathrm{T_k}$ of the IGM to $\mathrm{T_s}$ and this enables a large
fraction of the Universe to be visible in the brightness temperature
albeit with a few sources (top-left panel of the figure). The
brightness temperature does not go to zero at redshift six because, in
the model we have assumed, the Universe is not entirely ionized by
then (for recent papers on this issue see \cite{mesinger09}).

\subsection{Reionization histories: A comparison}
\label{subsec:reionhist}

In this section we compare the reionization histories for the three
scenarios explored above. To create a contigious observational cube or
``frequency cube'' ( right ascension (RA) $\times$ declination (DEC)
$\times$ redshift), we adopt the procedure described in
\cite{thomas09}. The procedure involves the stacking of RA and DEC
slices, taken from individual snapshots at different redshifts (or
frequency), interpolated smoothly to create a reionization
history. This data cube is then convolved with the point spread
function of the LOFAR telescope to simulated the mock data cube of the
redshifted 21-cm signal as seen by LOFAR. For further details on
creating this cube refer to \citet{thomas09}.

\begin{figure*}
\centering
\hspace{0cm}
\includegraphics[width=0.8\textwidth]{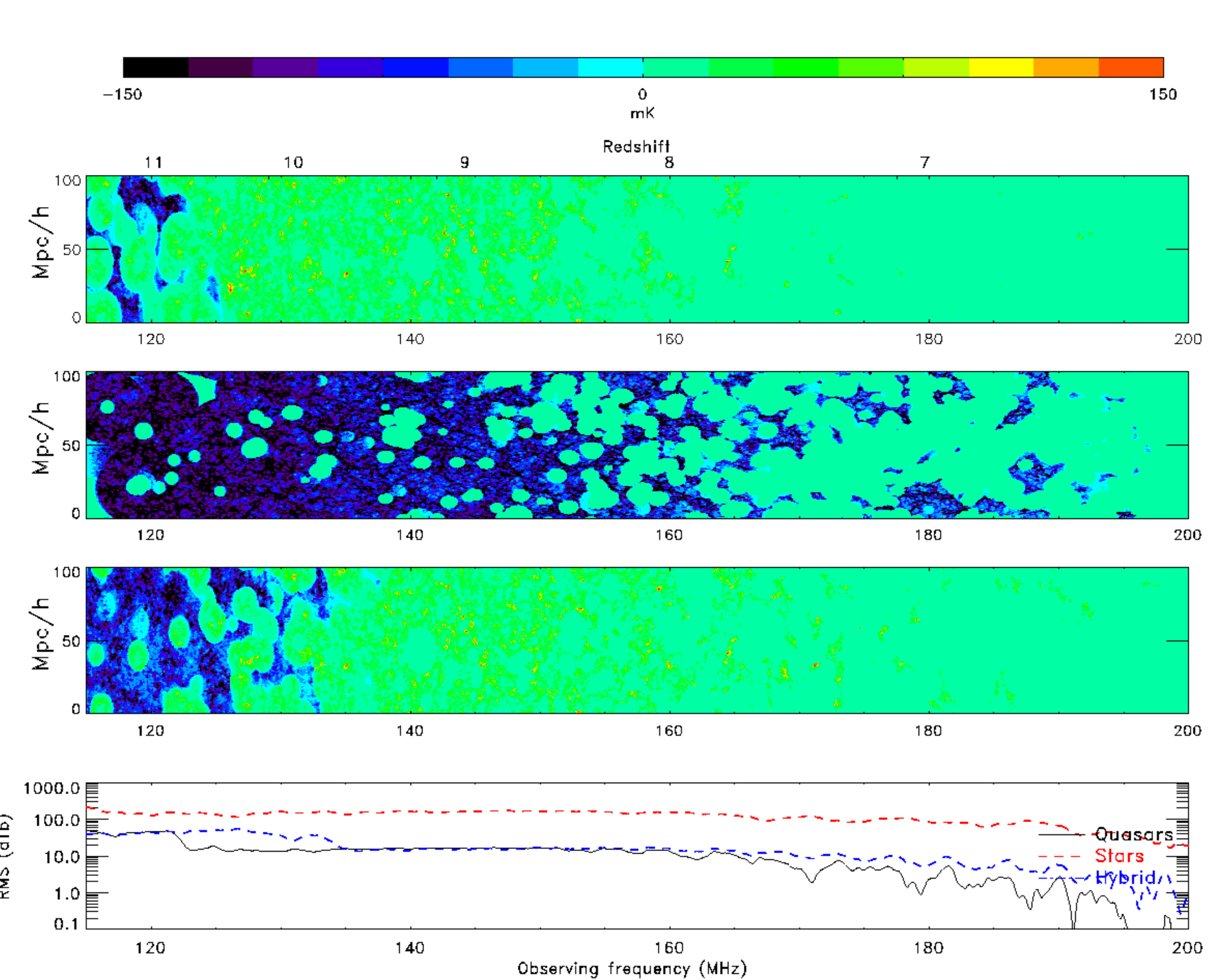}
\vspace{-.2cm}
\caption{\emph{Contrasting reionization histories}: From the top,
  reionization histories ($\delta \mathrm{T_b}$ in mK as a function of
  frequency or redshift) are plotted for miniqso, stellar and hybrid
  sources, respectively. The bottom panel plots the r.m.s of $\delta
  \mathrm{T_b}$ as a function of redshift/frequency for all the three
  cases.}
\label{fig:histcomp}
\end{figure*}

As expected the signatures (both visually and in the r.m.s) of the
three scenarios (Fig.~\ref{fig:histcomp}) are markedly different. In
the miniqso-only scenario, reionization proceeds extremely quickly and
the Universe is almost completely ($<\mathrm{x_{HII}}> = 0.95$) reionized
by around redshift 7. The case in which stars are the only source see
reionization end at a redshift of 6. Also in this case, compared to
the previous one, reionization proceeds in a rather gradual
manner. The hybrid model, as explained previously, interpolates
between the previous two scenarios.

\begin{figure*}
\centering
\hspace{0cm}
\includegraphics[width=0.8\textwidth]{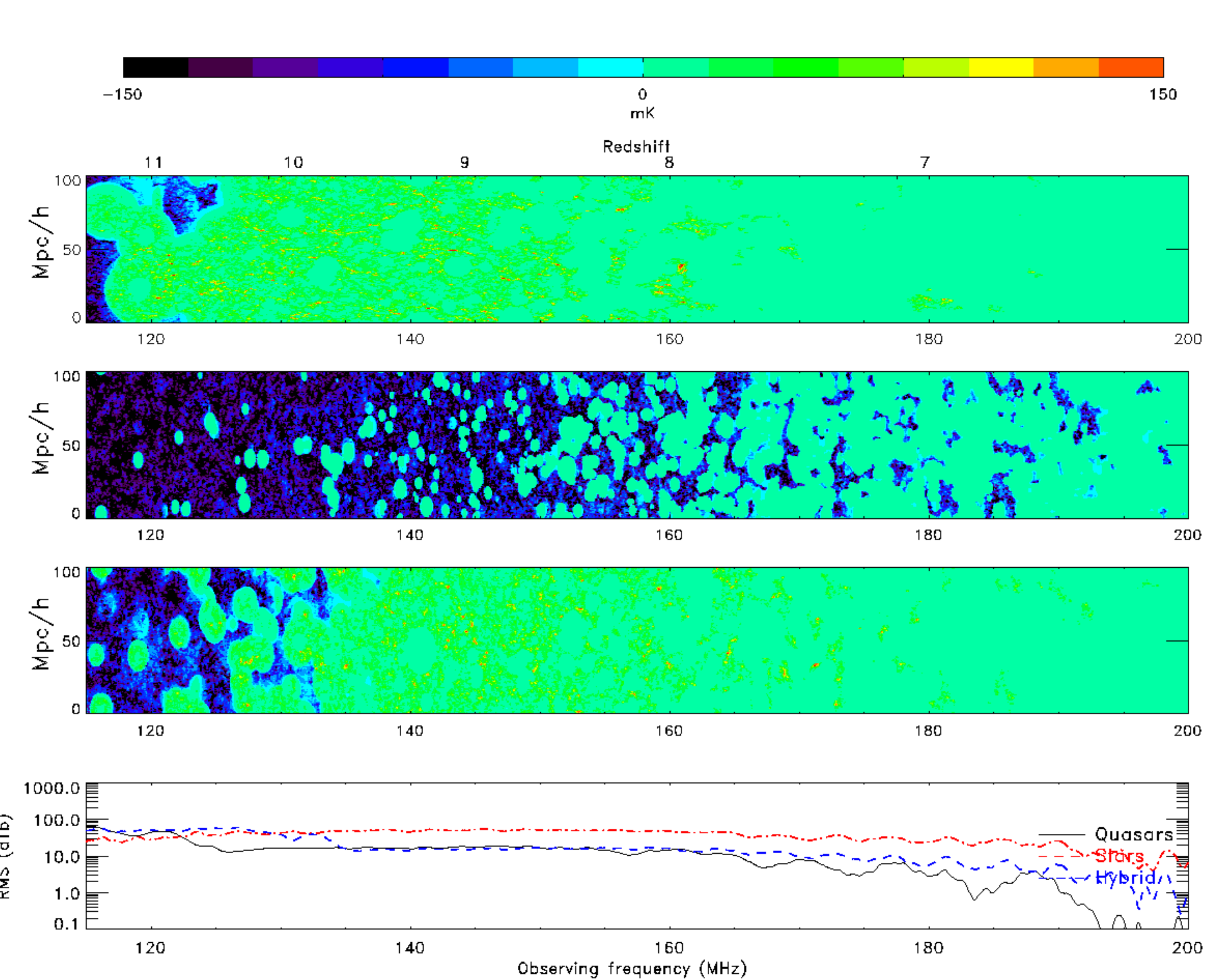}
\vspace{-.2cm}
\caption{\emph{Reionization histories assuming $\mathrm{T_s} = \mathrm{T_{k}}$}:
Reionization histories ($\delta \mathrm{T_b}$), same as in
Fig.~\ref{fig:histcomp} are plotted, now with the assumption that
the $\mathrm{T_s}$ has been completely decoupled from $\mathrm{T_{CMB}}$.
}
\label{fig:histcompinf}
\end{figure*}

The $\delta \mathrm{T_b}$ in Fig.~\ref{fig:histcomp} is calculated
based on the effectivness of $\mathrm{J_o}$, produced by the source, to
decouple the CMB temperature ($\mathrm{T_{CMB}}$) from the spin
temperature ($\mathrm{T_s}$). This flux, both in spatial extent and
amplitude, is obviously much larger in the case of miniqsos compared
to that of stars resulting in a markedly higher brightness
temperatures in both the miniqso-only and hybrid models when compared
to that of the stars. However, we know that stars themselves produce
Ly$\alpha$ radiation in their spectrum. Apart from providing
sufficient Ly$\alpha$ flux to their immediate surrounding, this
radiation builds up, as the Universe evolves, into a strong background
$\mathrm{J_o}$ \citep{ciardi03}, potentially filling the Universe with
sufficient Ly$\alpha$ photons to couple the spin temperature to the
kinetic temperature everywhere. Thus, we plot in
Fig.~\ref{fig:histcompinf} the same set of reionization histories as
in Fig.~\ref{fig:histcomp} but now assuming that $\mathrm{T_s}$
coupled to $\mathrm{T_k}$. Clearly the ``cold'' regions in the
Universe show up as regions with large negative brightness
temperature. This assumption though is not strictly applicable towards
the beginning of reionization. In \S\ref{sec:stats} we quantify the
differences in the brightness temperature evolution when this
assumption is made.

It has to be noted that the results we are discussing here are
extremely model dependent and any changes to the parametres can
influence the results significantly. This on the other is the
demonstration of the capability and the need for a ``\textsc{bears} like''
algorithm to span the enormous parametre space of the astrophysical
unknowns in reionization studies.

\subsection{Looking through LOFAR}
 To test the feasibiliy of 21-cm telescopes
(specifically LOFAR) in distinguishing observational signatures of
different sources of reionization, we need to propagate the
cosmological 21-cm maps generated in \S\ref{subsec:reionhist} using
LOFAR's telescope response. The point spread function (PSF) of the
LOFAR array was constructed according to the latest configuration of
the antenna layouts for the LOFAR-EoR experiment.

The LOFAR telescope will consist of up to 48 stations in The
Netherlands \footnote{This is decision under review and might change
  subsequently.} of which approximately 22 will be located in the core
region \citep{panos09}. The core marks an area of 1.7
$\times$ 2.3 kilometres and is essentially the most important part of
the telescope for an EoR experiment. The core station consists of two
sets of antennae, the Low Band (LBA; 30-90 MHz) and the High Band
Antenna (HBA; 110-240 MHz). The HBA antennae in the core stations is
further split into two ``half-stations'' of half the collecting area
(35-metre diameter), separated by $\approx$ 130 metres. This split
further improves the uv-coverage. For details on the antenna layout
and the synthesis of the antenna beam pattern refer to upcoming paper
by \citet{panos09}.

\begin{figure*}
\centering
\hspace{0cm}
\includegraphics[width=0.8\textwidth]{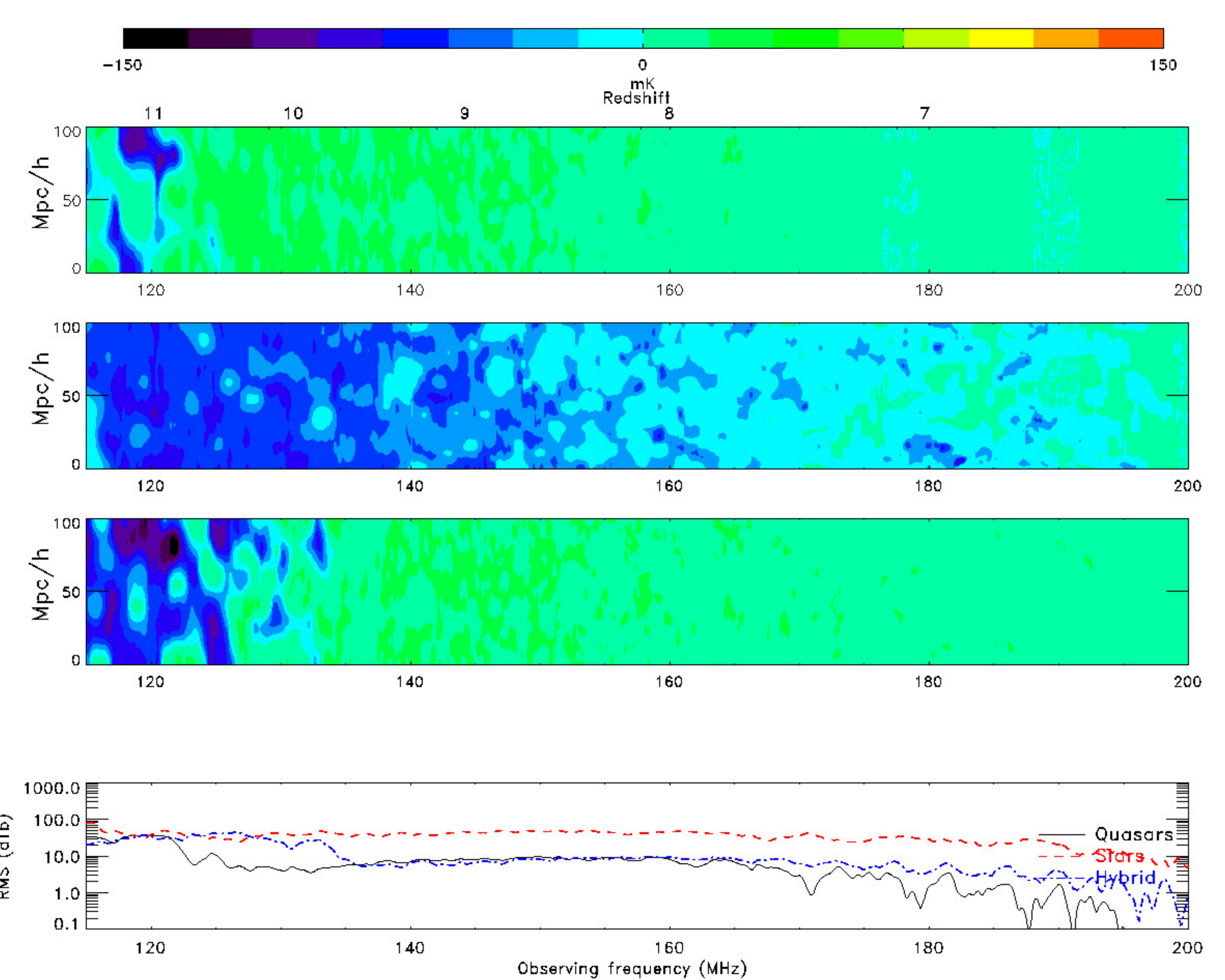}
\vspace{-.2cm}
\caption{\emph{Convolution with LOFAR - PSF}: Differential brightness
  temperatures after convolution with a point spread function
  corresponding to LOFAR is plotted for the same case as in
  Fig.~\ref{fig:histcomp}, with the temperatures plotted in mK.}
\label{fig:histcompafterconv}
\end{figure*}

In its current configuration, the resolution of the LOFAR-core is
expected to be around 3 arcmins. Thus, as an example at redshift 10,
all scales below $\approx$ 800 kpc will be filtered out.
Fig.~\ref{fig:histcompafterconv} shows this effect for the
reionization histories corresponding to Fig.~\ref{fig:histcomp}.  The
corresponding changes in the r.m.s of the brightness temperature is
also plotted. We see that, although smoothed to a large extent, there
still exists qualitative differences between the different
scenarios. This of course is a reflection of the vastly contrasting
models of reionization compared in this paper and difference will
disappear if the models investigated are more similar to each another.

\section{Statistical analysis of the simulations}
\label{sec:stats}

The previous section provided a qualitative description of the
differences in the 21-cm $\delta \mathrm{T_b}$ fluctuations for two
cases i.e., with and without the assumption that $\mathrm{T_s =
  T_k}$. In this section, a number of statistical tests are performed
to quantify these differences. For this purpose we focus on the
scenario with the miniqsos. This exercise can be repeated for the
other two scenarios as well.

\begin{figure}
\centering
\hspace{0cm}
\includegraphics[width=0.5\textwidth]{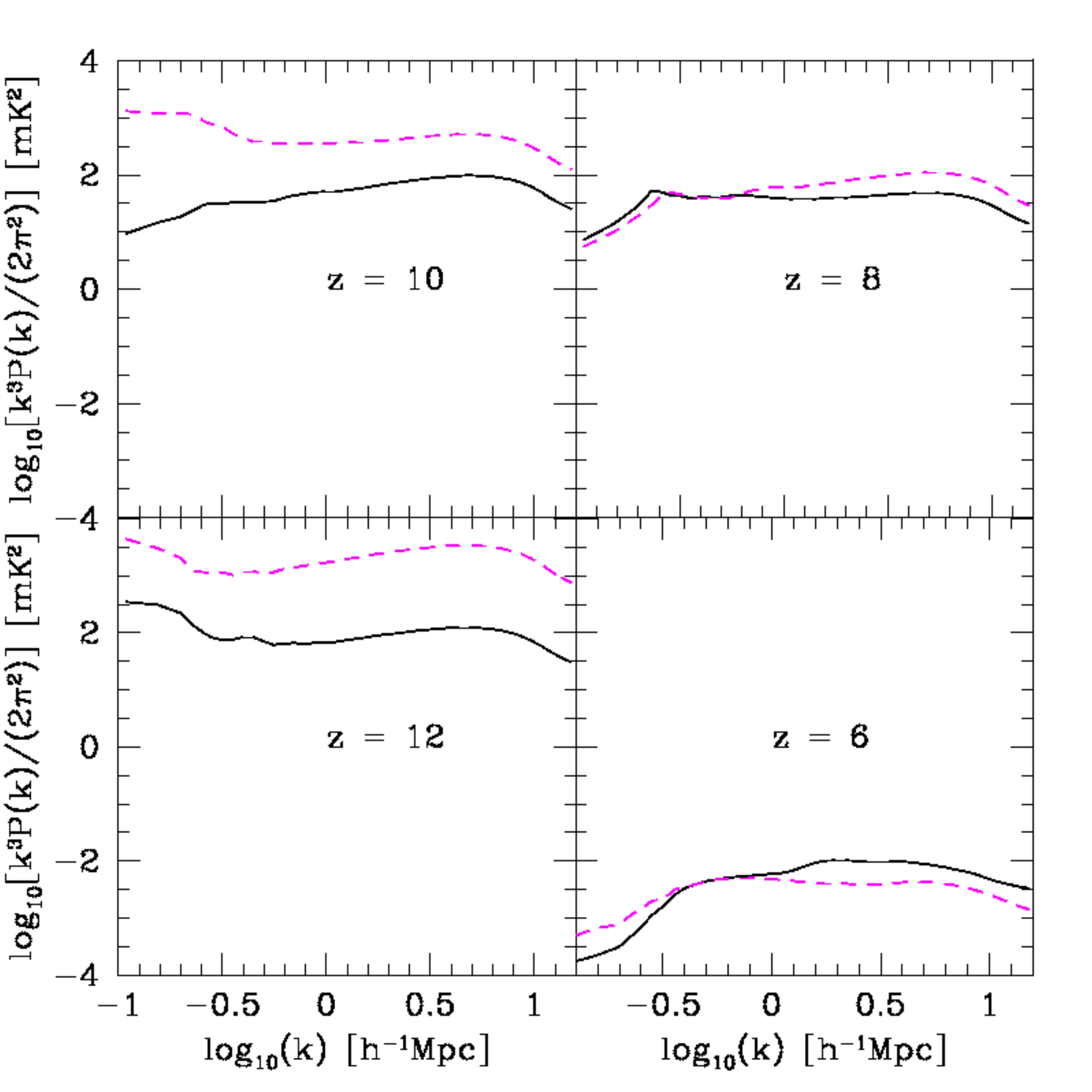}
\vspace{-.5cm}
\caption{The 3-D power spectra of $\delta \mathrm{T_b}$ at four
different redshifts (as indicated on the panels) are plotted. The
solid curves indicate the power spectra resulting from the
self-consistent inclusion of $\mathrm{T_s}$ and the dashed curve,
under the assumption $\mathrm{T_s} = \mathrm{T_k}$. At the onset of
reionization, $z \approx 12 $ in our model, $\delta \mathrm{T_b}$ is
larger at all scales under this assumption because we have enforced,
in effect, an \emph{artificial} decoupling of $\mathrm{T_s}$ from
$\mathrm{T_{CMB}}$ in regions of the simulated Universe where
$\mathrm{T_k} << \mathrm{T_{CMB}}$. But by redshift 10 the power
spectra match exactly as the Ly$\alpha$ becomes efficient everywhere
in the box to couple $\mathrm{T_s}$ to $ \mathrm{T_k}$. Notice that
the power keeps dropping as $\mathrm{x_{HI}} \rightarrow 0$ and
Universe gets reionized at lower redshifts.}
\label{fig:powspec}
\end{figure}

Firstly, Fig.~\ref{fig:powspec} shows the 3-D power spectra of $\delta
\mathrm{T_b}$ at redshift 12, 10, 8 and 6. In our model, reionization
begins at around redshift 12.7 and thus z=12 can be considered the
onset of reionization. The dashed curve reflects the assumption that
$\mathrm{T_s} = \mathrm{T_k}$ while the solid curve takes into account
the evolution of $\mathrm{T_s}$ self-consistently. At the initial
stages of reionization, only regions at close proximity to the source
have been ionized and heated, and the volume filling by ``spheres of
ionization'' has just begun, leaving large portions of the Universe
neutral and cooler than the CMB. Here we assume that after the epoch
of recombination the Universe cools adiabatically, i.e., $\mathrm{T_k}
\propto 1/(1+z)^2$ and $\mathrm{T_{CMB}} \propto 1/(1+z)$, which
results in $\mathrm{T_k} << \mathrm{T_{CMB}}$ at the onset of
reionization. Thus, we have a situation in which we have assumed
$\mathrm{T_s} = \mathrm{T_k}$, with $\mathrm{T_k} <<
\mathrm{T_{CMB}}$, resulting in an \emph{artificial} decoupling of
$\mathrm{T_s}$ and $\mathrm{T_{CMB}}$, and is manifested as increased
21-cm power. The discrepancy quickly (by $z = 10$) disappears as the
``spheres of Ly$\alpha$ `` overlap to volume-fill the simulation box
and hence increase the $\mathrm{T_s}$ --- $\mathrm{T_k}$ coupling
efficiency. Therefore, in our model, it is safe to make the assumption
that $\mathrm{T_s}$ follows $\mathrm{T_k}$ after about $z = 10$.

\begin{figure}
\centering
\hspace{0cm}
\includegraphics[width=0.5\textwidth]{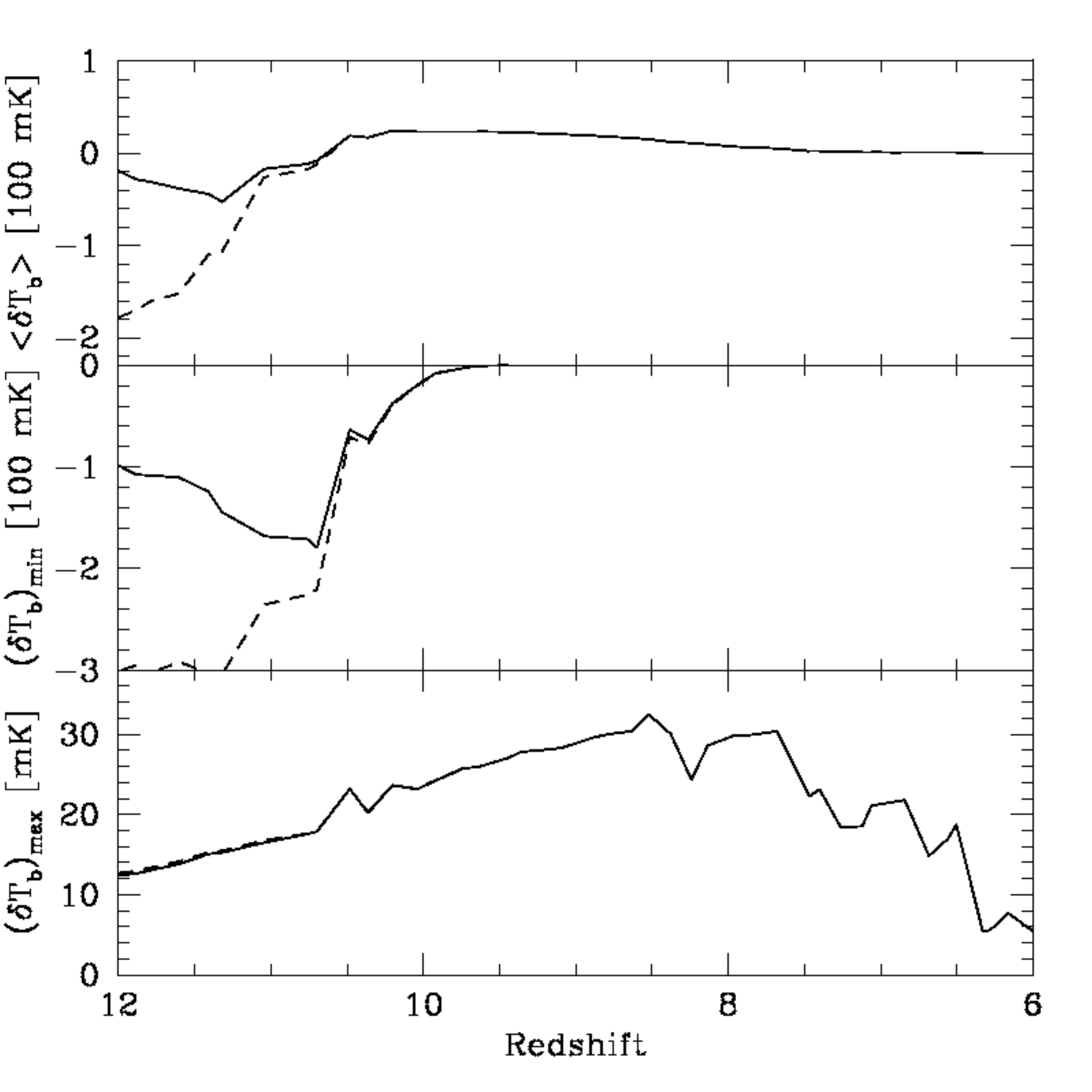}
\vspace{-.5cm}
\caption{Plotted above are the mean (top), minimum (middle) and
maximum (bottom) values of $\delta \mathrm{T_b}$ as a function of
redshift. Line styles correspond to Fig.\ref{fig:powspec}. Because at
earlier redshifts (around 12) $\mathrm{T_k} << \mathrm{T_{CMB}}$
within a large fraction in the simulation box, the minimum value
attained by assuming $\mathrm{T_s} = \mathrm{T_k}$ is much lower than when only
regions with Ly$\alpha$ excitation (which is typically near the
source) is allowed to couple $\mathrm{T_s}$ with $\mathrm{T_k}$. And
for the same reason the maximum value in both cases are similar
because the assumption holds close to the sources as the presence of
Ly$\alpha$ flux ensures efficient coupling. As a result the global
average $\delta \mathrm{T_b}$ is also lower for the approximated model
at the onset of reionization and catches up quickly to the ``proper''
model as the Universe fills up with sufficient Ly$\alpha$ flux.}
\label{fig:dtbmmm}
\end{figure}

From Eq.~\ref{eq:dtb} it follows that $\delta \mathrm{T_b}$ attains
large negative values when two conditions are satisfied,
$\mathrm{T_s} << \mathrm{T_{CMB}}$ and a high neutral hydrogen
overdensity.  We therefore plot the mean, minimum and maximum
values of $\delta \mathrm{T_b}$ as a function of redshift in
Fig. \ref{fig:dtbmmm}. As expected the mean and minimum values of
$\delta \mathrm{T_b}$ are much lower for the case in which the spin
temperature has been set to the kinetic temperature, because the
Universe has cooled substantially below the CMB and coupling
$\mathrm{T_k}$ everywhere in the Universe to $\mathrm{T_s}$ enforces
the conditions required for large negative $\delta \mathrm{T_b}$ as
stated above.  The maximum positive values though do not differ
because a positive $\delta \mathrm{T_b}$ indicates that $\mathrm{T_s}
> \mathrm{T_{CMB}}$ and hence has to be heated by the radiating
source. And in regions where the source has heated the IGM, the
presence of Ly$\alpha$ flux and colliosional coupling almost always
guarantees the coupling of $\mathrm{T_s}$ to $\mathrm{T_k}$ anyway.

\begin{figure}
\centering
\hspace{0cm}
\includegraphics[width=0.5\textwidth]{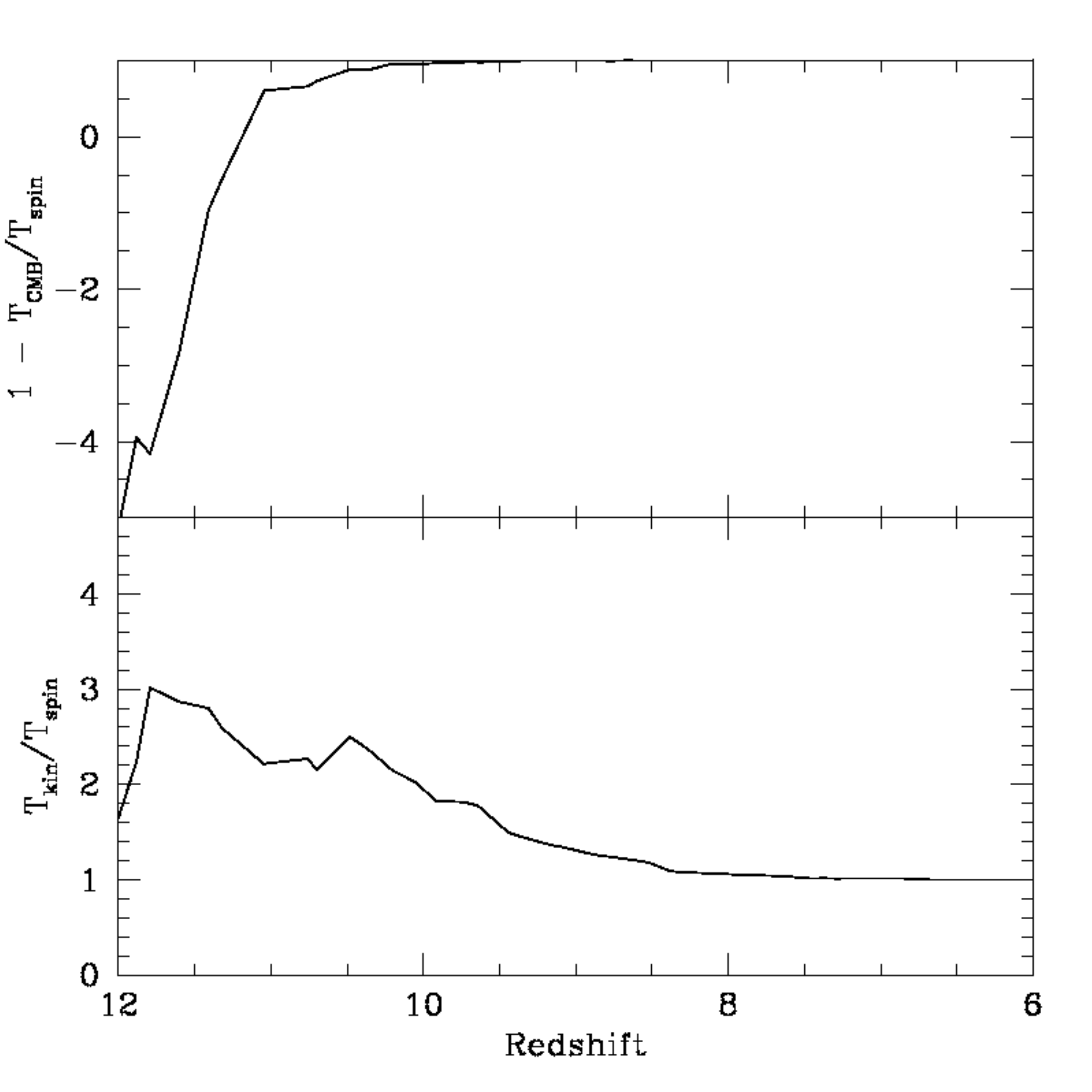}
\vspace{-.5cm}
\caption{ The differential temperature $\mathrm{(T_{spin} -
    T_{CMB})/T_{spin}}$ (top) and the ratio,
  $\mathrm{T_{kin}/T_{spin}}$ (bottom), are plotted as a function of
  redshift. If $\mathrm{T_{spin}} $ was assumed to track $\mathrm{T_{kin}}$, then the
  top panel would always be saturated at unity.  The kinetic
  temperature likewise merge with the spin temperature only by a
  redshift of $z = 8$, as seen in the bottom panel.}
\label{fig:meantstk}
\end{figure}

As another simple diagnostic of the efficiency of $\mathrm{T_s}$ ---
$\mathrm{T_k}$ coupling we plot, (i) the differential temperature,
i.e., $1 - \mathrm{T_{CMB}/T_{s}}$, (ii) their ratio,
  $\mathrm{T_k}/\mathrm{T_s}$, as a function of redshift
  (Fig. \ref{fig:meantstk}). Note here that if the spin temperature was
  artificially set to the kinetic temperature, the differential
  temperature would be saturated at unity. But the coupling ensures
  that the spin temperature remains below that of the CMB till about a
  redshift of 10.  Although significantly different at the early
  phases of reionization, in the model we have assumed, $\mathrm{T_s}$
  catches up with $\mathrm{T_k}$ at a redshift of about eight, when
  the coupling due to Ly$\alpha$ and collisions become significant
  enough to tie $\mathrm{T_s}$ to the local $\mathrm{T_k}$.

\begin{figure}
\centering
\hspace{0cm}
\includegraphics[width=0.5\textwidth]{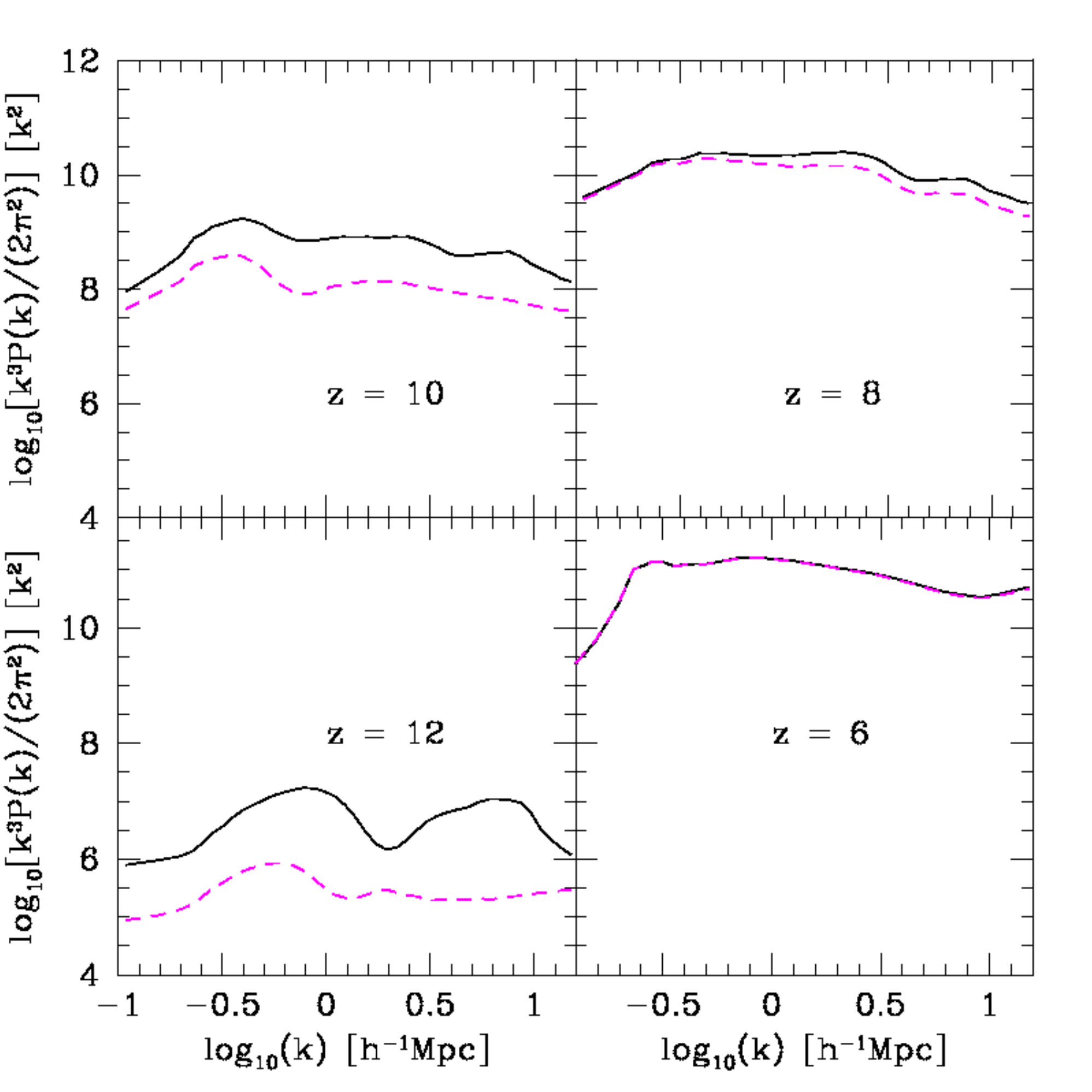}
\vspace{-.5cm}
\caption{At four different redshifts are plotted the power spectra of
the kinetic temperature (solid) and the spin temperature (dashed)
within the simulation box. Note how the $\mathrm{T_s}$ is considerably lower
than the $\mathrm{T_k}$ initially and then merges with $\mathrm{T_k}$ as reionization
proceeds. Refer text for more details.}
\label{fig:powspectemp}
\end{figure}

As an alternative probe to investigate the influence of
Ly$\alpha$-coupling coupling, in Fig.\ref{fig:powspectemp} we plot the
power spectra, now of the kinetic (solid line) and the spin (dashed
line) temperatures at four different redshifts. There are a few
interesting attributes to this figure. Firstly we see that at all
scales $\mathrm{T_s}$ is considerably lower than $\mathrm{T_k}$ at the
onset of reionization ($z \approx 12$). This scenario changes rapidly
as $\mathrm{T_s}$ approaches $\mathrm{T_k}$. Note that it is only
towards the end of reionization (z=6) that $\mathrm{T_s}$ is identical
to $\mathrm{T_k}$ at all scales. The pecularity observed is that all
scales in $\mathrm{T_s}$ do not approach $\mathrm{T_k}$
simultaneously. As seen from the top two panels of the figure at
$z=10$ and $z=8$, $\mathrm{T_s}$ approaches $\mathrm{T_k}$ at large
scales (small $k$ values). And then subsequently at small scales
towards the end of reionization. This is because the central parts of
the ionized bubble around the source was highly ionized and because
$J_o \propto \mathrm{x_{HI}}$, the spin temperature obtains extremely
low values. Now, as you go further away from the source, the IGM is
becomes neutral, but $\mathrm{J_o}$ from the source and the secondary
Ly$\alpha$ photons is high enough to couple the spin temperature to
the kinetic temperature. Thus, the large scales (large volumes outside
the ionized bubble) get ``matched'' first followed by the small
scales, when the IGM is hot and collisions become important.

\subsection{ Comparison to similar work}
\begin{figure}
\centering
\hspace{-.5cm}
\includegraphics[width=0.5\textwidth]{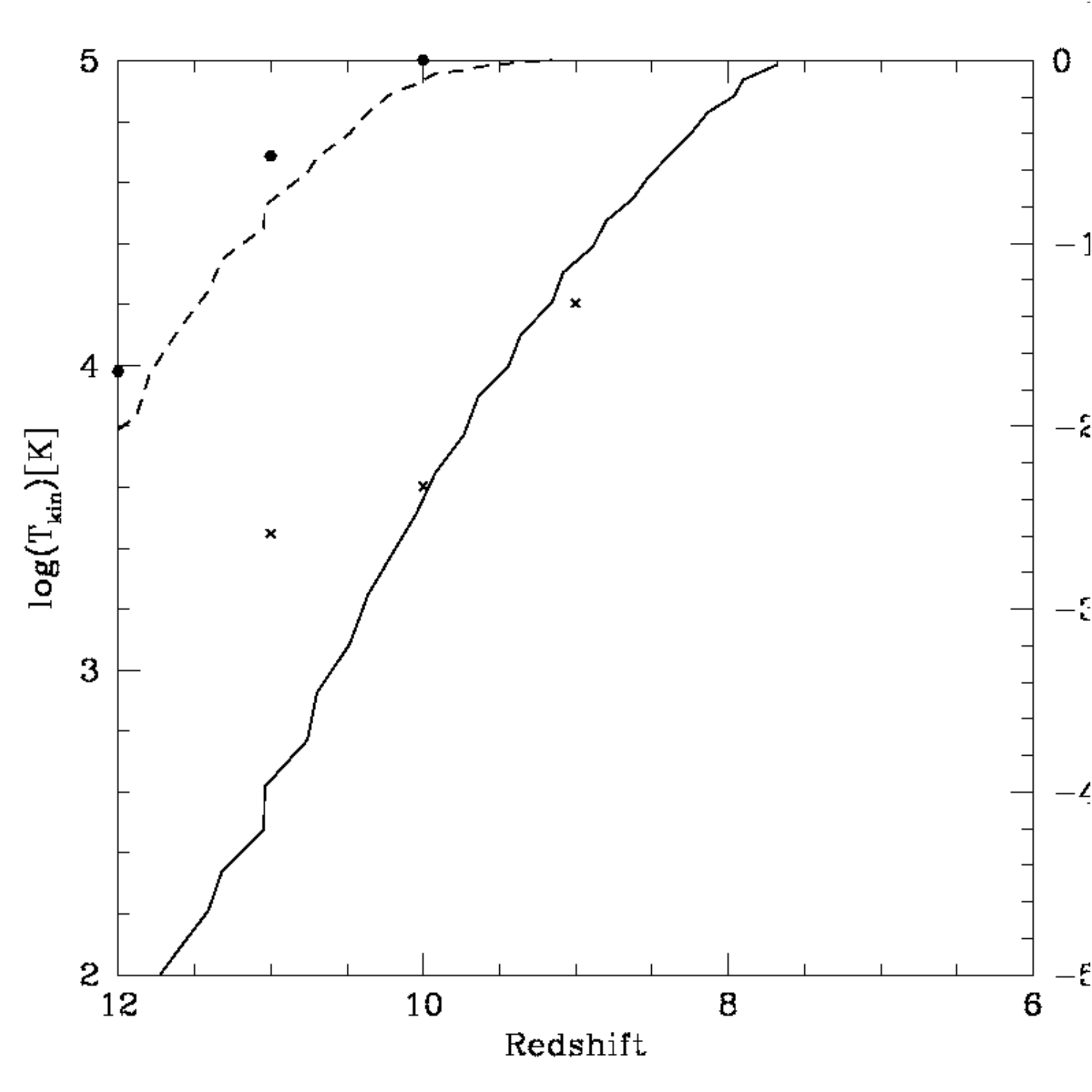}
\vspace{-.8cm}
\caption{ Plotted here are the evolution of the gas temperature
  $\mathrm{T_k}$ (solid line) and the volume filling factor (dashed
  line) of the X-ray radiation as a function of redshift for the case
  of the miniqsos. For comparison, points from \citet{santos08} are
  shown for $\mathrm{T_k}$ (crosses) and filling factor (filled
  hexagons) with their case of $f_x = 10$ and photon energies 100.0
  eV.}
\label{fig:compare}
\end{figure}

Here we briefly compare the output of our simulation with that of
\citet{santos08}, wherein a semi-analytical scheme is developed to
look at ionization and heating during the epoch of
reionization. Fig.\ref{fig:compare} shows the comparison between our
simulations which has now been performed for the case of a power-law
source starting at 100eV to facilitate a balanced comparison with the
100eV case of \citet{santos08}. Plotted in this figure are the average
gas temperature (solid line and crosses) and volume filling factor of
X-ray radiation (dashed line and filled hexagons). The lines
correspond to our simulation and points to \citet{santos08}. Their
case with the normalization factor $f_x =
10$, \footnote{\citet{santos08} normalized their spectrum assuming a
  total integrated power density to be $3.4 \times 10^{40} f_x$
  ergs/s/$Mpc^3$.} and photon energies of 100.0 eV shows similar
trends. Because we start reionization a bit later ($ z = 12 $) the
temperature is a bit lower initially.  
  Their model is qualitatively different because the X-ray emissivity
  is tied to their star-formation rate. Otherwise, the manner in which
  heating is implemented to similar to the one described in our case
  albeit without terms involving compton heating/cooling. These
  effects though are important towards the central regions of the
  source \citep{thomas08}.

\section{Conclusions \& Outlook}
\label{sec:conclusion}

The focus of this paper was three-fold. 1) to introduce an extension
to \textsc{bears} in order to incorporate heating (including X-ray
heating) and discuss its application to two cases of reionization and
heating, i.e., for blackbody (stellar) and power-law (miniqso) type
sources, 2) to monitor the evolution of spin temperature
($\mathrm{T_s}$) self-consistently and contrast its influence against
the usual assumption made in most reionization simulation that
$\mathrm{T_s} >> \mathrm{T_{CMB}}$, and 3) use the extended
\textsc{bears} code to study reionization due to a hybrid population
of stars and miniqsos as reionizing sources.

To incorporate heating into \textsc{bears}, we embeded spheres of
``kinetic temperature'' around the sources of radiation, much like in
the algorithm used to obtain the ionized fraction \citep{thomas09}.
The \textsc{bears} algorithm implemented is extremely fast, in that it
takes $\sim 5$ hours (2GHz, 16-core, 32GB shared RAM ) to perform the
radiative transfer (including heating) on about 35 different boxes
($512^3$ and 100~$h^{-1}$~comoving Mpc) between redshifts 12 and
6. These snapshots are interpolated between redshifts to produce a
contiguous data cube spanning redshifts $z \approx 6 - 12$, i.e., the
observational range of the LOFAR-EoR experiment. The efficiency of the
code facilitates the simulation of many different scenarios of
reionization and test their observational signatures. Apart from
predicting the nature of the underlying cosmological signal, these
simulations can be used in conjunction with others probes of
reionization to enhance the detectability and/or constrain parameters
concerning reionization. In \citet{jelic09} these simulations were
cross-correlated with the simulations of the CMB anisotropy (mainly
the kinetic and thermal SZ-effect) to probe reionization.  In the same
paper Jelic et al., repeated the same exercise by using reionization
simulations from \citet{mellema06} and found similar results, further
emphasizing the validity of the approximate manner in which we solve
the evolution of the kinetic temperature in our simulations.

The improved \textsc{bears} code was used to simulate three distinct
reionization scenarios, i.e., miniqsos, stellar and a mixture of
both. These simulations further emphasized that miniqsos were not only
very efficient in increasing $\mathrm{T_k}$ of the IGM but the
secondary Ly$\alpha$ flux produced by these sources was enough to
drive $\mathrm{T_s}$ away from $\mathrm{T_{CMB}}$ in an extended
region around the source thereby rendering the IGM around the source
``bright'' in $\delta \mathrm{T_b}$. This implies, if a miniqso
hosting a black hole in the mass range $> 10^7 ~M_\odot$ is within the
observing window of LOFAR, the value of $\delta \mathrm{T_b}$ and its
spatial extent would be conducive to a possible detection. The code
was also applied to simulate reionization with a hybrid population of
stars and miniqsos.  Every dark matter halo was embedded with both a
quasar and a stellar component. The relation between the two was set
according to the ``Magorrian relation'' we observe today. The total
mass in quasars were constrained following black hole mass density
estimates by \cite{volonteri08}. The effect of the hybrid population
on the IGM was an interpolation between the scenarios of stars and
miniqsos i.e., although the heating (and ionization) was not as
extended as in the case of miniqsos, it was definitely larger than
that for stars.

Our ultimate goal is to simulate mock data sets for the LOFAR-EoR
experiment. Towards this end, reionization simulations performed on 35
different snapshots between redshifts 6 and 12 were combined to form
``frequency cubes'' (reionization histories, \cite{thomas09}) for all
three scenarios i.e., stars, miniqsos and the hybrid case. To fold-in
the effect of the instrument we smoothed these frequency cubes
according to the LOFAR beam-scale. Notwithstanding the convolution by
the LOFAR beam, stark differences between the scenarios clearly
persists. The r.m.s $\delta \mathrm{T_b}$ cleary shows this difference
between scenarios.

All three scenarios of reionization discussed above was performed for
two different cases, i) assuming that $\mathrm{T_s}$ is always coupled
to $\mathrm{T_k}$ and ii) following the evolution of $\mathrm{T_s}$
self-consistently in the simulation. Figures \ref{fig:histcomp} and
\ref{fig:histcompinf} reflect the differences between the two
cases. To further quantify the influence of the self-consistent
evolution of $\mathrm{T_s}$, especially at the beginning of
reionization, a suite of statistical analyses was performed \footnote{
  Although these tests are specific to the scenario with miniqsos, it
  can be repeated for the case of stars and the hybrid. Miniqsos were
  chosen since in this case we observe the largest difference due to
  of the copous amounts of secondary Ly$\alpha$ flux and heating
  induced by miniqsos.}. Firstly, the 3-D power spectra of $\delta
\mathrm{T_b}$ (Fig. \ref{fig:powspec}) for the two cases was
calculated at four redshifts (12, 10, 8 and 6). We see that although
there is a considerable difference at the onset (z=12) of reionization
the power spectra of the brightness temperature match-up by a redshift
of 8. This is attributed to the volume filling nature of the spin
temperature coupling induced by miniqsos. The minimum, maximum and
mean value of $\mathrm{\delta T_b}$ as a function of redshift reveal
similar differences between the two cases. The maximum value of
$\mathrm{\delta T_b}$ is the same in both cases because regions where
the IGM is considerably heated also corresponds to regions with a
large supply of Ly$\alpha$ photons and they are also regions where
colliosional coupling is effective. The minimum and average values
reveal a difference since under the assumption that $\mathrm{T_s} =
\mathrm{T_{k}}$, regions much below $\mathrm{T_{CMB}}$ are
artificially coupled to $\mathrm{T_s}$. As a final statistic we
compute the 3-D power spectra of $\mathrm{T_k}$ and $\mathrm{T_s}$ and
compare them (Fig. \ref{fig:powspectemp}. Vast differences do exist
between the two almost till the end of reionization, when they become
identical.  An interesting result of this comparison is that the large
scales of the two spectra get matched before the small scales in the
box. The interpretation of this behaviour is given in section
\ref{sec:stats}. We also compared our results to that of
\cite{santos08} and found good match for the evolution of gas
temperature and the volume filling factor of X-ray radiation.

In the near future calibration errors expected from a LOFAR-EoR
observational run will be added to these simulations along with the
effects of the ionosphere and expected radio frequency interference
(RFI). Subsequently, Galactic and extra-galactic foregrounds modeled
as in \cite{jelic08} will be merged with these simulations to create
the final ``realistic'' data cube. These data cubes will be processed
using the LOFAR signal extraction and calibration schemes being
developed to retrieve the underlying cosmological signal in
preparation for the actual LOFAR-EoR experiment which has already
begun.

\section{Acknowledgment}
The author RMT would also like to thank the Institute for the Physics
and Mathematics of the Universe, Tokyo, for the support provided
during an extended stay.  LOFAR is being partially funded by the
European Union, European Regional Development Fund, and by
``Samenwerkingsverband Noord-Nederland'', EZ/KOMPAS.  The authors
would like to thank the anonymous referee was his/her extensive
comments on the manuscript and for improving its contents
significantly. RMT and SZ would also like to thank Joop Schaye and
Andreas H. Pawlik for providing us with the N-body simulations and for
useful discussions.

\label{lastpage}

\begin{thebibliography}{99}
\bibitem[\protect\citeauthoryear{Abel \& Norman}{2002}]{abel02}
 Abel T., Bryan G.~L., Norman M.~L., 2002, Science, 295, 93
\bibitem[\protect\citeauthoryear{Abel \& Norman}{2000}]{abel00}
 Abel T., Bryan G.~L., Norman M.~L., 2000, ApJ, 540, 39
\bibitem[\protect\citeauthoryear{Baek et al.}{2009}]{baek09} 
 Baek S., Di Matteo P., Semelin B., Combes F., Revaz Y., 2009, A\&A, 495, 389
\bibitem[\protect\citeauthoryear{Bromm et al.}{2002}]{bromm02} 
 Bromm, V., Coppi, P.S., Larson, R.B., 2002, ApJ, 564, 23
\bibitem[\protect\citeauthoryear{Chuzhoy, Alvarez, \&
 Shapiro}{2006}]{chuzhoy06} Chuzhoy L., Alvarez M.~A., Shapiro P.~R.,
 2006, ApJ, 648, L1
\bibitem[\protect\citeauthoryear{Ciardi \& Madau}{2003}]{ciardi03} 
 Ciardi B., Madau P., 2003, ApJ, 596, 1
\bibitem[\protect\citeauthoryear{Ciardi et al.}{2001}]{ciardi01} 
 Ciardi B., Ferrara A., Marri S., Raimondo G., 2001, MNRAS, 324, 381
\bibitem[\protect\citeauthoryear{Davis et al.}{1985}]{davis85} 
 Davis M., Efstathiou G., Frenk C.~S., White S.~D.~M., 1985, ApJ, 292, 371
\bibitem[\protect\citeauthoryear{Dijkstra, Haiman, \&
 Loeb}{2004}]{dijkstra04a} Dijkstra M., Haiman Z., Loeb A., 2004, ApJ,
 613, 646
\bibitem[\protect\citeauthoryear{Elvis et al.}{1994}]{elvis94} 
 Elvis M., et al., 1994, ApJS, 95, 1
\bibitem[\protect\citeauthoryear{Gnedin \& Abel}{2001}]{gnedin01} 
 Gnedin N.~Y., Abel T., 2001, NewA, 6, 437
\bibitem[\protect\citeauthoryear{Furlanetto, Zaldarriaga, \&
 Hernquist}{2004}] {furlanetto04b} Furlanetto S.~R., Zaldarriaga M.,
 Hernquist L., 2004, ApJ, 613, 1
\bibitem[\protect\citeauthoryear{Fan et al.}{2006}]{fan06}
 Fan X., et al., 2006, AJ, 131, 1203
\bibitem[\protect\citeauthoryear{Ferrarese}{2002}]{ferrarese02} 
 Ferrarese L., 2002, ApJ, 578, 90
\bibitem[\protect\citeauthoryear{Field}{1958}]{field58} 
 Field G.~B., 1958, Proc I.R.E., 46, 240
\bibitem[\protect\citeauthoryear{Fukugita \& Kawasaki}{1994}]{fuku94} 
 Fukugita M., Kawasaki M., 1994, MNRAS, 269, 563
\bibitem[\protect\citeauthoryear{Furlanetto \& Loeb}{2002}]{furlanetto02} 
 Furlanetto S.~R., Loeb A., 2002, ApJ, 579, 1
\bibitem[\protect\citeauthoryear{Jelic et al.}{2009}]{jelic09} 
 Jeli{\'c} V., et al., 2009, arXiv, arXiv:0907.5179 
\bibitem[\protect\citeauthoryear{Jeli{\'c} et al.}{2008}]{jelic08} 
 Jeli{\'c} V., et al., 2008, MNRAS, 891
\bibitem[\protect\citeauthoryear{Kuhlen \& Madau}{2005}]{kuhlen05} 
 Kuhlen M., Madau P., 2005, MNRAS, 363, 1069
\bibitem[\protect\citeauthoryear{Laor et al.}{1997}]{laor97} 
 Laor A., Fiore F., Elvis M., Wilkes B.~J., McDowell J.~C., 1997, ApJ, 477, 93 
\bibitem[\protect\citeauthoryear{Labropoulos et 
al.}{2009}]{panos09} Labropoulos P., et al., 2009, arXiv, 
arXiv:0901.3359 
\bibitem[\protect\citeauthoryear{Laursen et al.}{2009}]{laursen09} 
 Laursen P., Razoumov A.~O., Sommer-Larsen J., 2009, ApJ, 696, 853
\bibitem[\protect\citeauthoryear{Madau, Meiksin, \& Rees}{1997}]{madau97} 
 Madau P., Meiksin A., Rees M.~J., 1997, ApJ, 475, 429
\bibitem[\protect\citeauthoryear{Magorrian et al.}{1998}]{magorrian98} 
 Magorrian J., et al., 1998, AJ, 115, 2285 
\bibitem[\protect\citeauthoryear{Mellema et al.}{2006}]{mellema06} 
 Mellema G., Iliev I.~T., Alvarez M.~A., Shapiro P.~R., 2006, NewA, 11, 374
\bibitem[\protect\citeauthoryear{Mesinger}{2009}]{mesinger09} Mesinger A., 2009, arXiv, arXiv:0910.4161 
\bibitem[\protect\citeauthoryear{Mesinger \& Furlanetto}{2007}]{mesinger07} 
 Mesinger A., Furlanetto S., 2007, ApJ, 669, 663
\bibitem[\protect\citeauthoryear{Nakamoto, Umemura,\& Susa}{2001}]{nakamoto01} 
 Nakamoto T., Umemura M., Susa H., 2001, MNRAS, 321, 593
\bibitem[\protect\citeauthoryear{Nusser}{2005}]{nusser05} 
 Nusser A., 2005, MNRAS, 359, 183
\bibitem[\protect\citeauthoryear{Nusser \& Silk}{1993}]{nusser93} 
 Nusser A., Silk J., 1993, ApJ, 411, L1
\bibitem[\protect\citeauthoryear{Page et al.}{2007}]{page07} 
 Page L., et al., 2007, ApJS, 170, 335
\bibitem[\protect\citeauthoryear{Pawlik \& Schaye}{2008}]{pawlik08} 
 Pawlik A.~H., Schaye J., 2008, arXiv, 802, arXiv:0802.1715
\bibitem[\protect\citeauthoryear{Pierleoni et al.}{2009}]{pierleoni09} 
 Pierleoni M., Maselli A., Ciardi B., 2009, MNRAS, 393, 872 
\bibitem[\protect\citeauthoryear{Razoumov \& Cardall}{2005}]{razoumov05} 
 Razoumov A.~O., Cardall C.~Y., 2005, MNRAS, 362, 1413
\bibitem[\protect\citeauthoryear{Ricotti \& Ostriker}{2004}]{ricotti04a} 
 Ricotti M., Ostriker J.~P., 2004, MNRAS, 352, 547 
\bibitem[\protect\citeauthoryear{Rijkhorst et al.}{2006}]{rijkhorst06} 
 Rijkhorst E.-J., Plewa T., Dubey A., Mellema G., 2006, A\&A, 452, 907
\bibitem[\protect\citeauthoryear{Ritzerveld, Icke, \& Rijkhorst}{2003}]{ritzerveld03} 
 Ritzerveld J., Icke V., Rijkhorst E.-J., 2003, astro, arXiv:astro-ph/0312301
\bibitem[\protect\citeauthoryear{Santos et al.}{2008}]{santos08} 
Santos M.~G., Amblard A., Pritchard J., Trac H., Cen R., Cooray A., 2008, ApJ, 689, 1
\bibitem[\protect\citeauthoryear{Sazonov, Ostriker, \&
 Sunyaev}{2004}]{sazonov04} Sazonov S.~Y., Ostriker J.~P., Sunyaev
 R.~A., 2004, MNRAS, 347, 144
\bibitem[\protect\citeauthoryear{Schaerer}{2002}]{schaerer02} 
 Schaerer D., 2002, A\&A, 382, 28
\bibitem[\protect\citeauthoryear{Seljak \& Zaldarriaga}{1996}]{seljak96} 
 Seljak U., Zaldarriaga M., 1996, ApJ, 469, 437
\bibitem[\protect\citeauthoryear{Shull \& van Steenberg}{1985}]{shull85} 
 Shull J.~M., van Steenberg M.~E., 1985, ApJ, 298, 268
\bibitem[\protect\citeauthoryear{Spergel et al.}{2007}]{spergel07}
 Spergel D.~N., et al., 2007, ApJS, 170, 377
\bibitem[\protect\citeauthoryear{Springel}{2005}]{springel05} 
 Springel V., 2005, MNRAS, 364, 1105 
\bibitem[\protect\citeauthoryear{Springel \& Hernquist}{2003}]{springel03} 
 Springel V., Hernquist L., 2003, MNRAS, 339, 312 
\bibitem[\protect\citeauthoryear{Susa}{2006}]{susa06} 
 Susa H., 2006, PASJ, 58, 445
\bibitem[\protect\citeauthoryear{Thomas et al.}{2009}]{thomas09} 
 Thomas R.~M., et al., 2009, MNRAS, 393, 32
\bibitem[\protect\citeauthoryear{Thomas \& Zaroubi}{2008}]{thomas08} 
 Thomas R.~M., Zaroubi S., 2008, MNRAS, 384, 1080 
\bibitem[\protect\citeauthoryear{Tasitsiomi}{2006}]{tasitsiomi06} 
 Tasitsiomi A., 2006, ApJ, 645, 792
\bibitem[\protect\citeauthoryear{Vanden Berk et al.}{2001}]{vandenberk01} 
 Vanden Berk D.~E., et al., 2001, AJ, 122, 549
\bibitem[\protect\citeauthoryear{Vignali, Brandt, \&
 Schneider}{2003}]{vignali03} Vignali C., Brandt W.~N., Schneider
 D.~P., 2003, AJ, 125, 433
\bibitem[\protect\citeauthoryear{Volonteri, Lodato, \&
 Natarajan}{2008}] {volonteri08} Volonteri M., Lodato G., Natarajan
 P., 2008, MNRAS, 383, 1079
\bibitem[\protect\citeauthoryear{Vald{\'e}s \& Ferrara}{2008}]{valdes08} 
 Vald{\'e}s M., Ferrara A., 2008, MNRAS, 387, L8 
\bibitem[\protect\citeauthoryear{Whalen \& Norman}{2006}]{whalen06} 
 Whalen D., Norman M.~L., 2006, ApJS, 162, 281
\bibitem[\protect\citeauthoryear{Wouthuysen}{1952}]{wouthuysen52} 
 Wouthuysen S.~A., 1952, AJ, 57, 31
\bibitem[\protect\citeauthoryear{Wyithe \& Loeb}{2004}]{wyithe04} 
 Wyithe J.~S.~B., Loeb A., 2004, ApJ, 610, 117
\bibitem[\protect\citeauthoryear{Yoshida et al.}{2003}]{yoshida03} 
 Yoshida N., Sokasian A., Hernquist L., Springel V., 2003, ApJ, 598, 73
\bibitem[\protect\citeauthoryear{Zahn et al.}{2007}]{zahn07} 
 Zahn O., Lidz A., McQuinn M., Dutta S., Hernquist L., Zaldarriaga M., 
 Furlanetto S.~R., 2007, ApJ, 654, 12 
\bibitem[\protect\citeauthoryear{Zaroubi et al.}{2007}]{zaroubi07}
 Zaroubi S., Thomas R.~M., Sugiyama N., Silk J., 2007, MNRAS, 375, 1269 
\bibitem[\protect\citeauthoryear{Zaroubi \& Silk}{2005}]{zaroubi05} 
 Zaroubi S., Silk J., 2005, MNRAS, 360, L64
\end{thebibliography}
\end{document}